\documentclass[aps,nopacs,showkeys,nofootinbib,preprint]{revtex4}
\usepackage{epsfig,amssymb,bm,graphicx,xcolor,amsmath,rotating}
\usepackage[leftcaption]{sidecap}
\usepackage{setspace}
\usepackage{slashed}
\usepackage{amssymb}
\usepackage{tabularx}
\usepackage{tikz-cd}

\newcommand{\Nth}{N_{\mathrm{ctrl}}}
\newcommand{\uth}{u_{\mathrm{th}}}
\newcommand{\etaLCFA}{ \eta_{\mathrm{LCFA}}}

\newcommand{\eg}{e.g.\ }
\newcommand{\abs}[1]{\left\vert #1 \right\vert}

\begin{document}

\title{Note on Klein-Nishina effect in strong-field QED: \\
the case of nonlinear Compton scattering} 

\author{U. Hernandez Acosta${}^{1, 2}$, B.~K\"ampfer${}^{1, 3}$}
\affiliation{${}^{1}$Helmholtz-Zentrum  Dresden-Rossendorf, 01328 Dresden, Germany}
\affiliation{${}^{2}$Center for Advanced Systems Understanding (CASUS), Helmholtz-Zentrum
Dresden-Rossendorf, Untermarkt 20, 02826 G\"orlitz, Germany}
\affiliation{${}^{3}$Institut f\"ur Theoretische Physik, TU~Dresden, 01062 Dresden, Germany}

\date{\today}

 
\begin{abstract}
Suitably normalized differential probabilities of one-photon emission in external electromagnetic fields
are compared to quantify the transit of nonlinear Compton
scattering to linear Compton scattering, described by the Klein-Nishina formula, and to constant crossed field treatment.
The known Klein-Nishina suppression at large energies is further enforced by increasing field intensity.
In view of the Ritus-Narozhny conjecture, we demonstrate that different paths in the field intensity vs.\
energy plane towards large values of the quantum non-linearity parameter $\chi$ facilitate 
significantly different asymptotic dependencies, both in the Klein-Nishina regime and the constant crossed field regime
and in between.      
  \end{abstract}


\maketitle

\section{Introduction}

The Klein-Nishina effect means a dropping cross section of the linear Compton scattering \cite{Klein:1929lcc} from
Thomson's value at small energies towards high energies.
In astrophysics, this dropping reduces the energy loss of ultra-relativistic electrons and positrons \cite{Fang:2020dmi,Schlickeiser:2009qq}
and the inverse Compton scattering rates with ultra-high energy photons, most notable for energies in the order of PeV
\cite{LHAASO:2023gne,LHAASO:2021gok}.
Additionally, in certain sites, ultra-intense electromagnetic background fields are present and enable or influence elementary
charged-particle--photon interactions. Pulsars in general and magnetars in particular represent such sites.
In earth-bound laboratories similarly extreme conditions can be achieved, hence linking to or exploring 
analog astrophysics scenarios. Beam-beam interactions in accelerators \cite{Yakimenko:2018kih}
or beam-laser interactions \cite{Abramowicz:2021zja,Salgado:2021fgt,Storey:2023wlu}
or high-intensity laser-electron (plasma \cite{Brodin:2022dkd,Seipt:2023bcw,Zhang:2020lxl}) 
experiments \cite{Poder:2017dpw,Cole:2017zca} 
or laser-laser interactions \cite{Blinne:2018nbd,Gies:2017ezf,Gies:2017ygp}
are to be mentioned in that context.

A special branch of phenomenology is tied to strong-field Quantum Electro Dynamics (sfQED) 
(see \cite{Fedotov:2022ely} for a recent survey with an extensive citation list and
\cite{HernandezAcosta:2023msl} for a concise formulation in momentum space)
aiming at describing the fundamental electromagnetic interactions in or at a background field.
The latter one could be of electromagnetic
(or electro-weak) nature or being gravity or something - hypothetically - beyond the Standard Model of Particle Physics.
We confine ourselves to electromagnetic backgrounds, i.e. sfQED.\footnote{
Many entries on the topic are cited in \cite{Fedotov:2022ely}.
Further surveys are provided by \cite{Gonoskov:2021hwf,Titov:2015pre,DiPiazza:2011tq,Kaminski:2009wwd}.
}

A straightforward question is then whether a noticeable background field has a significant impact on 
elementary processes. We narrow here the problem to the impact of an electromagnetic field on the
photon emission off electrons and consider specifically Compton scattering.\footnote{
A few entries for Compton as one-photon emission are
\cite{Ritus85,Lotstedt:2009zz,Mackenroth:2010jr,Hartin:2011vr,
Heinzl:2009nd,Seipt:2010ya,Seipt:2011zz,Seipt:2011dx,Seipt:2013taa,Seipt:2016rtk,
Titov:2015tdz,Dinu:2018efz,Ilderton:2018nws,Ilderton:2020dhs,
Nielsen:2021nuo,Gelfer:2022nqy,Kampfer:2020cbx,Acosta:2021iyu,Titov:2023yev,Angioi:2016vir,Blackburn:2018sfn};
for two-photon Compton (two-vertex diagram) see, e.g., 
\cite{Herrmann:1971jc,Guccione-Gush:1975juz,Akhiezer:1985,Ehlotzky:1989,Mackenroth:2012rb,Seipt:2012tn,Dinu:2018efz,Seipt:2013hda},
and higher orders in \cite{Lotstedt:2012zz,Lotstedt:2013uya,Dinu:2019pau}.
}

One aspect concerns the
Klein-Nishina effect in a strong background field. Another issue is related to the Ritus-Narozhny conjecture,
arguing that the account of the background field within the Furry picture is restricted to not too intense fields
\cite{Ritus:1972ky,Fedotov:2016afw,Heinzl:2021mji,Mironov:2020gbi}.
In doing so we resort to fairly known formalism as a vehicle to elucidate, numerically, links between established descriptions
of linear Compton scattering, nonlinear Compton scattering, and photon emission in a constant crossed field.
In this context, we not only review well-known results but also extend the comparative analysis of these findings, which are, in part, widely dispersed across highly specialized publications.
This subject is also of interest for approximations implemented in simulation codes - both for astrophysics
applications and laser physics in particular (cf.\ 
\cite{DiPiazza:2018bfu,DiPiazza:2017raw,Gelfer:2022nqy,Nielsen:2021nuo,Blackburn:2021cuq,Dinu:2019pau,
Ilderton:2018nws,Heinzl:2020ynb}).

Our note is organized as follows. In section II, we recall formulas describing the normalized probabilities of 
linear and nonlinear Compton scattering as well as photon emission in a constant crossed field.
Subsequently, the dependence on a Lorentz invariant quantity characterizing the emitted photon (Section III), 
the energy dependence (Section IV) and external field intensity dependence (Section V) are analyzed.
Section VI demonstrates how various paths approach a large value of the so-called 
quantum non-linearity parameter $\ chi$ leaves a strong impact on the asymptotic. After a brief discussion in Section VII,
we conclude in Section VIII. A few relevant numbers are listed in Appendix \ref{app.A},
and Appendix \ref{app.B} recaps relations of plane wave and constant crossed field quantities. Appendix \ref{app.D} presents master curves of the harmonics structures. 
Appendix \ref{app.C} relates the invariant differential cross section and the photon-energy and photon-angular
differential cross sections.

\section{Exclusive one-photon emission: recollection of linear Compton vs.\ nonlinear Compton vs.\ constant crossed field}

\begin{figure}[h!]
\includegraphics[width=0.45\columnwidth]{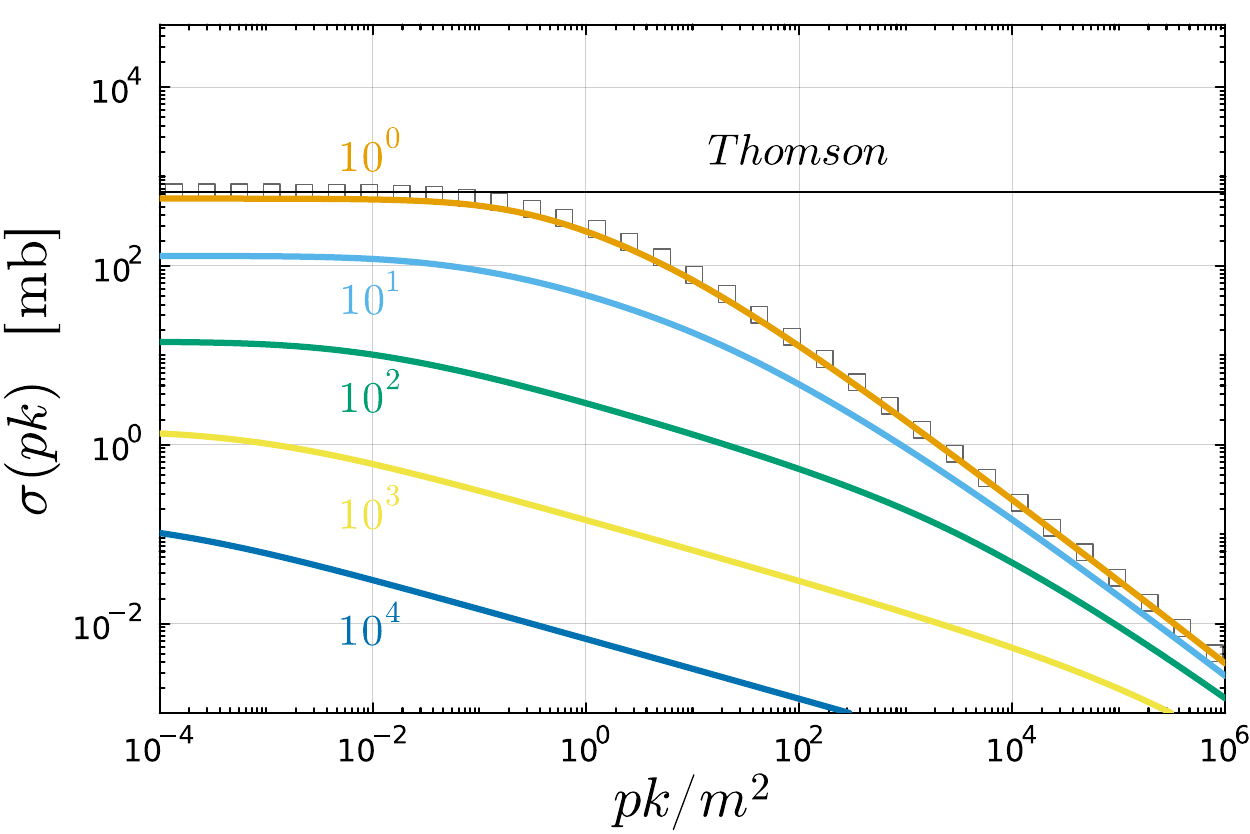} 
\includegraphics[width=0.45\columnwidth]{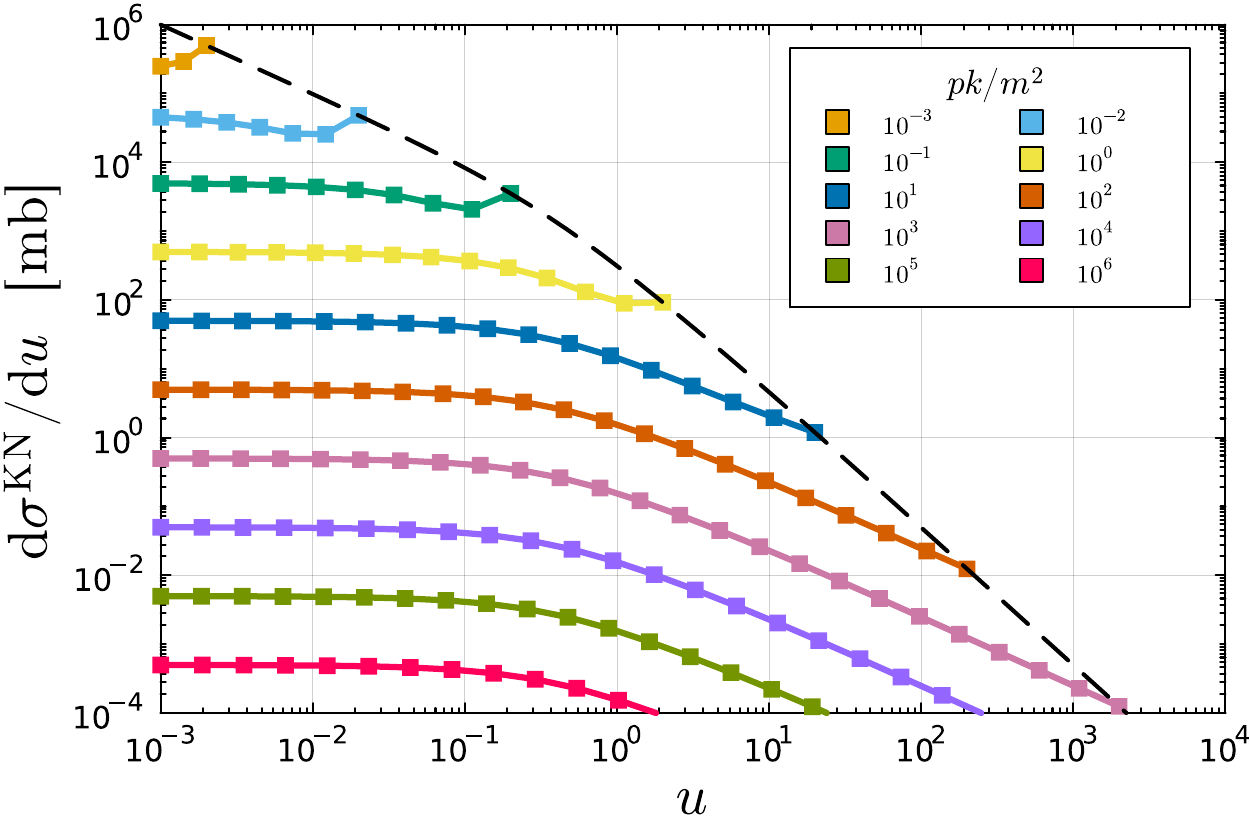} 
\caption{Left panel: Total KN cross section Eq.~(\ref{eq.1}) (squares) and the IPA cross section 
$\sigma^\mathrm{IPA} = \int_0^\infty d u \, \frac{d \sigma^\mathrm{IPA}}{du}$
for various values of the field intensity parameter $a_0 = 1$, 10, $10^2$, $10^3$ and $10^4$
(solid curves, from top to bottom).
Right panel: Differential KN cross section Eq.~(\ref{eq.10}) as a function of $u$ for various values of the energy.
The dashed curve connects the loci of kinematic limits $u = 2 p \cdot k/m^2$.
Given applications in high-intensity laser physics and high-energy astrophysics 
(cf.\ Appendix \ref{app.A}), we have displayed the KN spectra for a fairly large energy range.
\label{fig.1}
}
\end{figure}

The Klein-Nishina (KN) effect means a dropping cross section of the linear Compton scattering \cite{Klein:1929lcc} from
Thomson's value $\sigma^\mathrm{Thomson} = \frac{8 \pi}{3} r_e^2$ 
at small energies, $p \cdot k /m^2 = (s-m^2)/m^2 \to 0$, towards high energies:\footnote{
We employ units with
$\hbar = c = 1$ and note the electron radius of $r_e = \alpha /m$ with Sommerfeld's fine-structure constant
$\alpha \approx 1/137$ (ignoring for the sake of clarity the running coupling) and electron mass $m$;
the electron's and photon's $in$ four-momenta are $p$ and $k$, respectively 
(for $out$-states, the four-momenta receive primes). 
The Mandelstam variable $s = (p + k)^2$ refers to the notion of ``energy".
}  
\begin{align}
\sigma^\mathrm{KN} &= \frac{2 \pi r_e^2}{u_{KN}}
\left[ \left( 1 - \frac{4}{u_\mathrm{KN}} - \frac{8}{u_\mathrm{KN}^2} \right) \log(1+u_\mathrm{KN})
+ \frac12 + \frac{8}{u_\mathrm{KN}} - \frac{1}{2 (1+u_\mathrm{KN})^2} \right] \label{eq.1}\\[2mm]
&= \left\{   
\begin{array}{ll}
\frac{8 \pi r_e^2}{3} (1-u_\mathrm{KN} ) & \mathrm{for} \, u_\mathrm{KN} \ll 1\\
\pi r_e^2 \frac{1}{u_\mathrm{KN}} \left( 2 \log u_\mathrm{KN} + 1 \right) & \mathrm{for} \, u_\mathrm{KN} \gg 1 
\end{array}  \right. ,
\end{align}
where $u_\mathrm{KN} := 2 p \cdot k /m^2$, see empty squares in Fig.~\ref{fig.1}-left.
At the origin of such behavior is the differential cross section \cite{LLIV}
\begin{align}
\frac{d \sigma^\mathrm{KN}}{du} &= \frac{2 \pi \alpha^2}{p\cdot k} \frac{1}{(1+u)^2}
\left(1 + \underline{\frac{u^2}{2(1+u)}}
- 2 \frac{u}{u_{KN}} \left[1- \frac{u}{u_\mathrm{KN}} \right] \right) \label{eq.10}\\[2mm]
&= \frac{2 \pi \alpha^2}{p \cdot k} \times \left\{ \begin{array}{ll}
1 \,  - 2 u (1+ u_\mathrm{KN}^{-1}) + u^2 \left(\frac72 
\frac{4}{u_\mathrm{KN}} + \frac{2}{u_\mathrm{KN}^2} \right) + \mathcal{O}(u^3)
&\mathrm{for} \, u \to 0\\
\frac{2 + 2u_\mathrm{KN} + u_\mathrm{KN}^2}{2 (1 + u_\mathrm{KN})^3} 
\stackrel{u_\mathrm{KN} \gg 1}{\longrightarrow} 
\frac14 \frac{m^2}{p\cdot k} \, &\mathrm{for} \, u = u_\mathrm{KN}\\
\end{array}  \right. ,
\end{align}
where $u \in [0, u_{KN}]$ denotes the Ritus invariant  
$u = \frac{k\cdot k'}{k \cdot p'}$ \cite{Ritus85}
which is related, via energy-momentum balance $p + k = p' + k'$, 
to the lightfront momentum fraction of the $out$-photon 
$\mathfrak{s}= \frac{k \cdot k'}{k \cdot p}$ via $\mathfrak{s}= \frac{u}{1+u}$. 
The underlined term in Eq.~(\ref{eq.10}) gives rise to the logarithmic
high-energy behavior $\sigma^\mathrm{KN} \propto (p\cdot k/m^2)^{-1} \log (p \cdot k/m^2)$.
With increasing energy, the differential cross section drops at small values of $u$
as $\propto (p\cdot k/m^2)^{-1}$ which turns, for $u > 1$, into  $\propto (p\cdot k/m^2)^{-2}$,
see Fig.~\ref{fig.1}-right. Quite qualitatively, one can argue that, with increasing energy, more and more
exclusive and inclusive channels open with the consequence that one particular exclusive channel receives less
probability of keeping the unitarity of the S matrix.  
In the case of linear Compton scattering, often the electron recoils are said to reduce the 
$out$ photon energy.

The above (differential) cross section is for the leading-order tree-level term of exclusive
$2 \to 2$ scattering $e^- + \gamma \to {e^-}' + \gamma'$ within perturbative QED,
organized in powers of $\alpha$
(for higher orders, cf.\ \cite{Lee:2021iid}).
It is known  \cite{LLIV,Ritus85} to emerge from nonlinear Compton scattering,
$e_A^- \to {e_A^-}' + \gamma'$, as leading-order weak-field term, where $e_A^-$ is for the electron wave function
in an external, classical background field $A$ which may refer to a monochromatic laser field in a plane-wave model
as the so-called infinite plane-wave approximation (IPA).
Having this relation in mind, one can define a suitably normalized probability\footnote{The relation of Ritus' differential probability \cite{Ritus85} and the differential cross section is given by \cite{Ivanov:2004fi,Acosta:2021iyu} \( \mathrm{d}W/\mathrm{d}u = a_0^2 m^2 (p\cdot k)/(e^2 q^0) \, \mathrm{d}\sigma/\mathrm{d}u  \), where \( q^0 \) is the zeroth component of the effective momentum of the incoming electron.}
(henceforth dubbed "cross section")
which makes this relation explicit:\footnote{
See \cite{Bragin:2020akq} for the definition of ``local cross sections" for processes in an external field.
Instead of using wave packages, we employ a circularly polarized, monochromatic plane-wave field
as a proxy of a mono-energetic photon beam under special circumstances.
The intensity parameter is $a_0$.
}
\begin{align}
\frac{d \sigma^\mathrm{IPA}}{du} &=
\frac{2 \pi \alpha^2}{p\cdot k} \frac{1}{(1+u)^2} \sum_{n=1}^{N \to \infty}
\Theta (u_n - u) \mathcal{F}_n (z_n), \label{eq.5}\\
& \mathcal{F}_n = - \frac{2}{a_0^2} J_n^2 (z_n) 
+ A \left[J_{n+1}^2 (z_n) + J_{n-1}^2 (z_n) - 2 J_n^2 (z_n)  \right] \\
& z_n =\frac{2 na_0}{\sqrt{1 + a_0^2}} \sqrt{\frac{u}{u_n} \left(1 - \frac{u}{u_n} \right)} \label{eq.7}\\
& u_n = \frac{2 n}{1 + a_0^2} \frac{p\cdot k}{m^2} \quad
\to n \ge n_{min} \equiv \frac12 u (1+a_0^2) \frac{m^2}{p \cdot k} \\
& A = 1 + \frac{u^2}{2 (1+u)} \label{eq.8},
\end{align}
see \cite{Acosta:2021iyu,Kampfer:2020cbx} and \cite{Ritus85} for the basics.
The classical field strength parameter characterizes the background field
$a_0$, and the meaning of the variable $u$ is as above.
The differential cross section depends separately on two variables: 
the energy variable $p \cdot k/m^2$ and intensity parameter $a_0$,
which can be combined with the so-called ``quantum non-linearity" parameter $\chi = a_0 \, p\cdot k/m^2$ 
and a second one. 
The total cross section\footnote{
The subsequent numerical evaluations uncover a large range of numerical values of the involved parameters.
Therefore, the summation of a large number of harmonics,
$N = \mathcal{O}(10^9)$ in Eq.~(\ref{eq.5}), is often required. 
(Based on properties of the Bessel functions of large arguments and large indices,
harmonics up $N = \mathcal{O} (a_0^3)$ should be accounted for.)
In accomplishing that goal, we employ the series expansion method referred to as Richardson extrapolation \cite{Sidi:2003bnm}, which drastically improves the rate of convergence.} $\sigma^\mathrm{IPA}$ is compared with $\sigma^\mathrm{KN}$
in Fig.~\ref{fig.1}-left. In addition to the KN effect, the external field causes a further suppression,
increasing with $a_0$. In line with the above-quoted argument, one can expect that, according to unitarity.
In other words, the exclusive decay of an electron state, $e_A^- \to {e_A^-}' + \gamma'$,
quantified by the total cross section, becomes more and more reduced with increasing environmental field $A$.
The degree of suppression depends on the energy, at a given field strength $A$.
This is our executive summary, which is put subsequently into various perspectives
in a differential manner. As far as we know, this knowledge appears to be relatively uncommon. 

At large values of $a_0$, the differential probability behind Eq.~(\ref{eq.5}) turns in the expression for one-photon emission
in a constant crossed field (CCF).
Transferred to ``cross section", linked in an {\it ad hoc} manner to  Eq.~(\ref{eq.5}), 
with some seemingly abuse of relations of $\chi$ and $a_0$ and energy 
(see however \cite{King:2014wfa} for a rationale and discussion in Appendix \ref{app.B}), one has \footnote{\label{fn.6}The locally constant field approximation (LCFA) uses the rate
$\frac{d R_\mathrm{LCFA}}{d \mathfrak{s}} = 
\frac{d \sigma^\mathrm{CCF}_{a_0 \to a_0 g(\phi)}}{d \mathfrak{s}} \frac{a_0^2 g(\phi)^2}{4 \pi \alpha m^2}$,
related to CCF Eqs.~(\ref{eq.6} - \ref{eq.12}),
where $\mathfrak{s} = u/(1+u)$ is the above mentioned  lightfront momentum transfer fraction, 
see \cite{Fedotov:2022ely} for details.
The quantity $g(\phi)$ is the pulse envelope describing the pulse evolution as a function of the invariant phase $\phi$.
Analogously, the locally monochromatic approximation (LMA) is related to IPA via
$\frac{d R_\mathrm{LMA}}{d \mathfrak{s}} = \frac{d \sigma^\mathrm{IPA}_{a_0 \to a_0 g(\phi)}}{d \mathfrak{s}} 
\frac{a_0^2 g(\phi)^2}{4 \pi \alpha m^2}$, 
where in IPA Eqs.~(\ref{eq.5} - \ref{eq.8}) the replacement $a_0 \to a_0 g(\phi)$ also applies
and $u = s/(1-s)$ is to be used.
(Mimicking pulses by $a_0 \to a_0 g(\phi)$ has been traded, according to \cite{Heinzl:2020ynb}
for the first time, in \cite{Titov:2019kdk} in an {\sl ad hoc} manner.)
Both LCFA and LMA are often used approximation schemes, implemented in many simulation codes 
for laser-particle/plasma interactions. 
LMA is shown in \cite{Heinzl:2020ynb} to approach LCFA in the strong-field limit.
The rates $R_\mathrm{LCFA}$ and $R_\mathrm{LMA}$,
defined by Eqs.~(84) and (98) in \cite{Fedotov:2022ely}, require the $\mathfrak{s}$ and $\phi$ integrations,
see also Eqs.~(1, 2) in \cite{Ilderton:2018nws}.
For more details on LCFA and LMA and improvements, see \cite{DiPiazza:2018bfu,DiPiazza:2017raw,Ilderton:2018nws}.
\color{black}
}
\begin{align}
\frac{d \sigma^\mathrm{CCF}}{du} &= - \frac{4 \pi \alpha^2}{m^2 a_0} \frac{1}{\chi} \frac{1}{(1+u)^2} 
\mathcal{F}_\mathrm{CCF} (\hat z) ,  \label{eq.6}\\
& \mathcal{F}_\mathrm{CCF} (\hat z) = \frac{2}{\hat z} \left[1 + \frac{u^2}{2 (1+u)} \right] \Phi'(\hat z)
+ \int_{\hat z}^\infty dy \, \Phi(y) , \quad \Phi = \pi Ai \\
& \hat z = \left( \frac{u}{\chi} \right)^{2/3}, \quad 
\chi = a_0 \frac{p\cdot k}{m^2} \mathrm{~is~independent~of~} \omega,  
\label{eq.12}
\end{align}
which in turn obeys the asymptotic \cite{Ritus85}
\begin{align}
\frac{d \sigma_{\mathrm{large}-u}^\mathrm{CCF}}{du} =
\frac{2 \sqrt{\pi} \alpha^2}{m^2 a_0} \chi^{-1/2} u^{-3/2}
\exp\left\{ - \frac{2 u}{3 \chi} \right\}, \quad \mathrm{for~} u \gg 1, \chi .
\label{eq.9}
\end{align}

Such a discussion is important w.r.t.\ the Ritus-Narozhny conjecture \cite{Ritus:1972ky,Narozhnyi:1980dc}
(cf.\ \cite{Fedotov:2022ely}  for an up-to-date discussion with citations therein)
which argues that the underlying Furry picture\footnote{
Meaning here that the external field is exactly accounted for in the quasi-classical electron wave function.
For one-photon emission, the Furry-picture Feynman diagram reads
\setlength{\unitlength}{1mm}
\thicklines
\begin{picture}(8,5)
\put(0,0.5){\line(1,0){8}}
\put(0,1.0){\line(1,0){8}}
\put(4.0,0.5){\vector(0,1){5}}
\put(4,0.75){\circle*{1}}
\end{picture}
where the double line is for the wave function $\Psi$ solving the Dirac equation
$(i \slashed{\partial}_x - e \slashed{A} - m) \Psi(x) = 0$ for an electron (charge $e$)
in the external field $A(x)$.
}
breaks down at $\alpha \chi^{2/3} \ge 1$ and loop corrections become severe.
In fact, Refs.\ \cite{Ilderton:2019kqp,Podszus:2018hnz} point out that the large-$\chi$ regime
may be achieved by large energy or high intensity which both have very distinctive asymptotic, i.e.\
$\chi$ alone is clearly not sufficient to characterize the process.
Besides the principal importance of such considerations, they are also tightly related to the design
of approximation schemes implemented in simulation codes.

Given that specific framework, we quantify the relations of the above-quoted cross sections on a differential level
and point out the detailed conditions for $d \sigma^\mathrm{IPA}/du  \to d \sigma^\mathrm{CCF}/du$ and
$d \sigma^\mathrm{IPA}/du  \to d \sigma^\mathrm{KN}/du$. We also show that approaching large values
of $\chi$ along different paths in the $a_0$-$p\cdot k/m^2$ plane facilitates different asymptotic.  

\section{The $\mathbf{u}$ dependence}

\begin{figure}[h!]
\includegraphics[width=0.7\textwidth,trim=3.0cm 0.0 0.0 0.0]{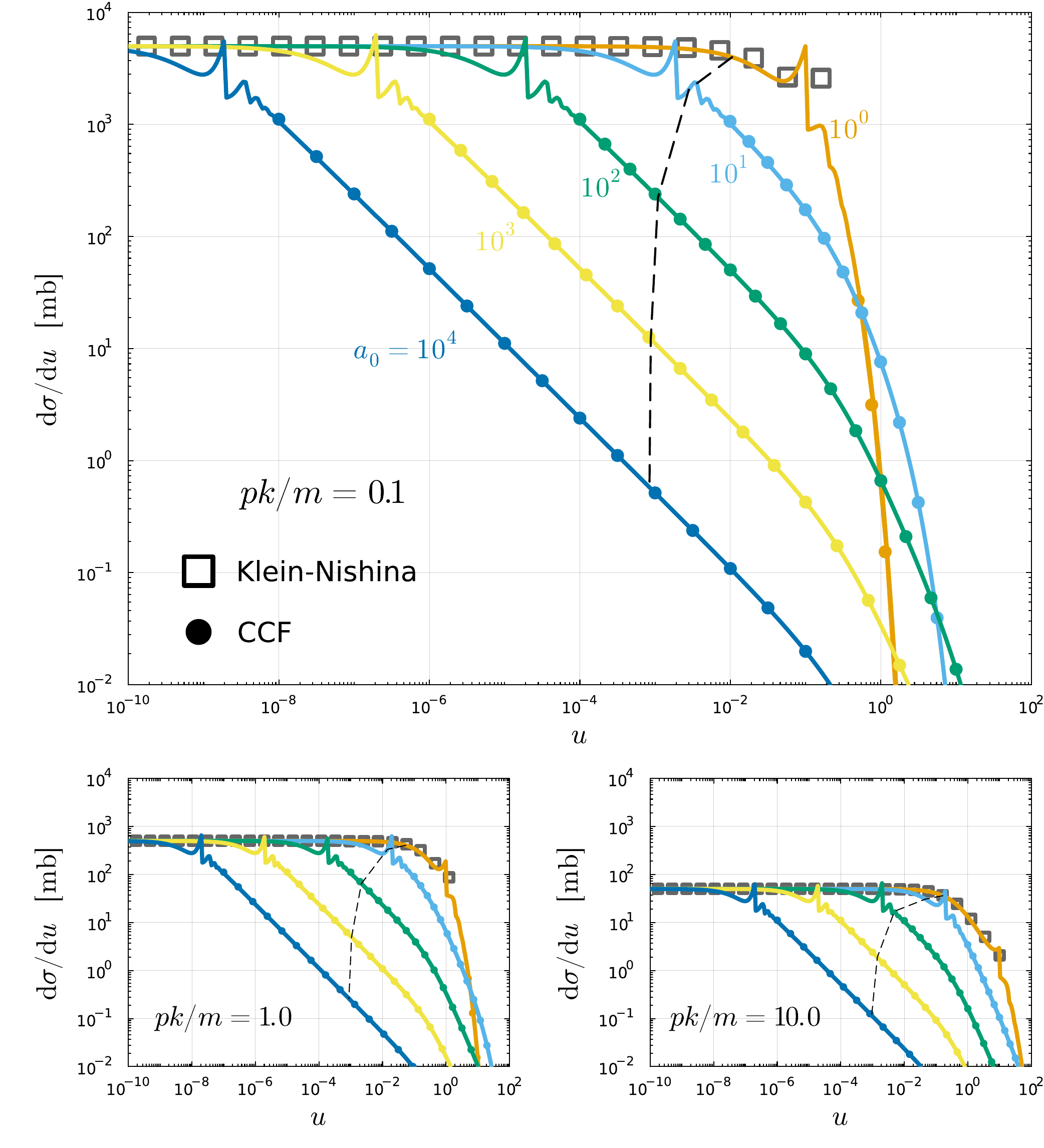} 
\caption{Differential cross section as a function of $u$ for $p\cdot k/m^2 = 0.1$ (top panel), 1 (left bottom panel)
and 10 (right bottom panel). IPA, Eq.~(\ref{eq.5}),
for various values of $a_0 = 1, 10^1, 10^2, 10^3$ and $10^4$
(from top to bottom), is exhibited by solid curves.
The squares are for KN, Eq.~(\ref{eq.10}).
Circles depict CCF, Eq.~(\ref{eq.6}),
beyond the harmonic structures.
(At smaller values of $u$, CCF fails badly by a strong overshot, even diverging for $u \to 0$.)
Right/below the thin dashed curves, 90\% of the total cross sections is located.
\label{fig.a0dep}
}
\end{figure}

Figure \ref{fig.a0dep} exhibits the differential cross section $d \sigma^\mathrm{IPA} / du$
as a function of $u$ for various values of $a_0 = 10^0 \cdots 10^3$. Squares display the KN cross section, while the CCF cross section is marked by dots.
The left panel is for $p \cdot k /m^2 = 10^{-1}$, while the right panel has  $p \cdot k /m^2 = 10^1$.
With increasing values of $a_0$ the ``spectra" are shifted to the left, as also do the harmonic structures.
The behavior in the IR region, i.e.\ at small $u$, is determined by
\begin{align}
\frac{d \sigma^\mathrm{IPA}}{du}\bigg|_{u \to 0} \approx
\frac{2 \pi \alpha^2}{p\cdot k} \left\{ 1 - u \left[2 + \frac{1 + 2 a_0^2}{p\cdot k/m^2} \right]\right\}
 \approx \frac{d \sigma^\mathrm{KN}}{du}\bigg|_{\mathrm{small~}u} - \frac{2 \pi \alpha^2}{p\cdot k}
\, u \, \frac{2 a_0^2}{p\cdot k/m^2}, \label{eq.14}
\end{align}
and obeys the $a_0$ independent limit
$\lim_{u \to 0} \frac{d \sigma^\mathrm{IPA}}{du}  = \frac{d \sigma^\mathrm{KN}}{du}  
= \frac{2 \pi \alpha^2}{p\cdot k}$ as already emphasized in \cite{DiPiazza:2018bfu,DiPiazza:2017raw} .
The last term in Eq.~(\ref{eq.14}) quantifies the $a_0$ dependence.
That is, upon integration, the total IPA cross section contains inevitably a KN-like contribution.
To quantify that we exhibit by the thin dashed curve the demarcation
where 10\% (on left/above) vs.\ 90\% (on right/below) of the total IPA cross section is located which clearly
stems from the large-$u$ sections, despite the small differential cross section.\footnote{
In doing so we inspect the partially integrated cross section
$\sigma (U) := \int_0^U du \, \frac{d \sigma^\mathrm{IPA}}{du}$,
normalized to the total cross section, which resembles a $\tanh$ function. 
}
At sufficiently large values of $u$, the IPA differential cross section is identical to the CCF expression
(marked by dotted curves), remarkably also for $a_0 = \mathcal{O} (1)$. 
The overall patterns are the same, irrespective of whether 
the energy is small (as in the left panel, where $\chi = 10^{-1} \cdots 10^2$
at $p\cdot k/m^2 = 10^{-1}$) 
or large (as in right panel, where $\chi = 10^1 \cdots 10^4$ at $p\cdot k/m^2 = 10^1$ ).

As an interim summary, we mention:
$d \sigma^\mathrm{IPA} / d u \stackrel{\mathrm{small}~u}{ \xrightarrow{\hspace*{0.9cm}}}  
d \sigma^\mathrm{KN} / d u$,
$d \sigma^\mathrm{IPA} /d u \stackrel{\mathrm{large}~u}{ \xrightarrow{\hspace*{0.9cm}}}  
d \sigma^\mathrm{CCF} /d u$
when keeping fixed $a_0$ and $p\cdot k/m^2$;
the notions ``small"/``large" depend separately on $a_0$ and $p\cdot k/m^2$.
The harmonic structure at $u = u_1$ may serve as demarcation via
\begin{align}
\frac{d \sigma^\mathrm{IPA}}{d u}\bigg|_{u = u_1} = 
\frac{2 \pi \alpha^2}{p \cdot k} \frac{1}{(1 + u_1)^2} \left(1 + \frac{u_1^2}{2(1+u_1)} \right)
\approx \frac{d \sigma^\mathrm{KN}}{d u} \bigg|_{u = 0} \times
\left\{ \begin{array}{ll}
1 & \mathrm{for~} u_1 \ll 1 , \\
\frac{1 + a_0^2}{4} \frac{m^2}{p \cdot k} & \mathrm{for~ }u_1 \gg 1 , \\
\end{array} \right.
\end{align}
where $u_1 \stackrel{\gg}{\ll} 1$ corresponds to $\frac{p \cdot k}{m^2} \stackrel{\ll}{\gg} 1 + a_0^2$. 

\section{The energy dependence}

\begin{figure}[h!]
\includegraphics[width=0.7\textwidth,trim=3.0cm 0.0 0.0 0.0]{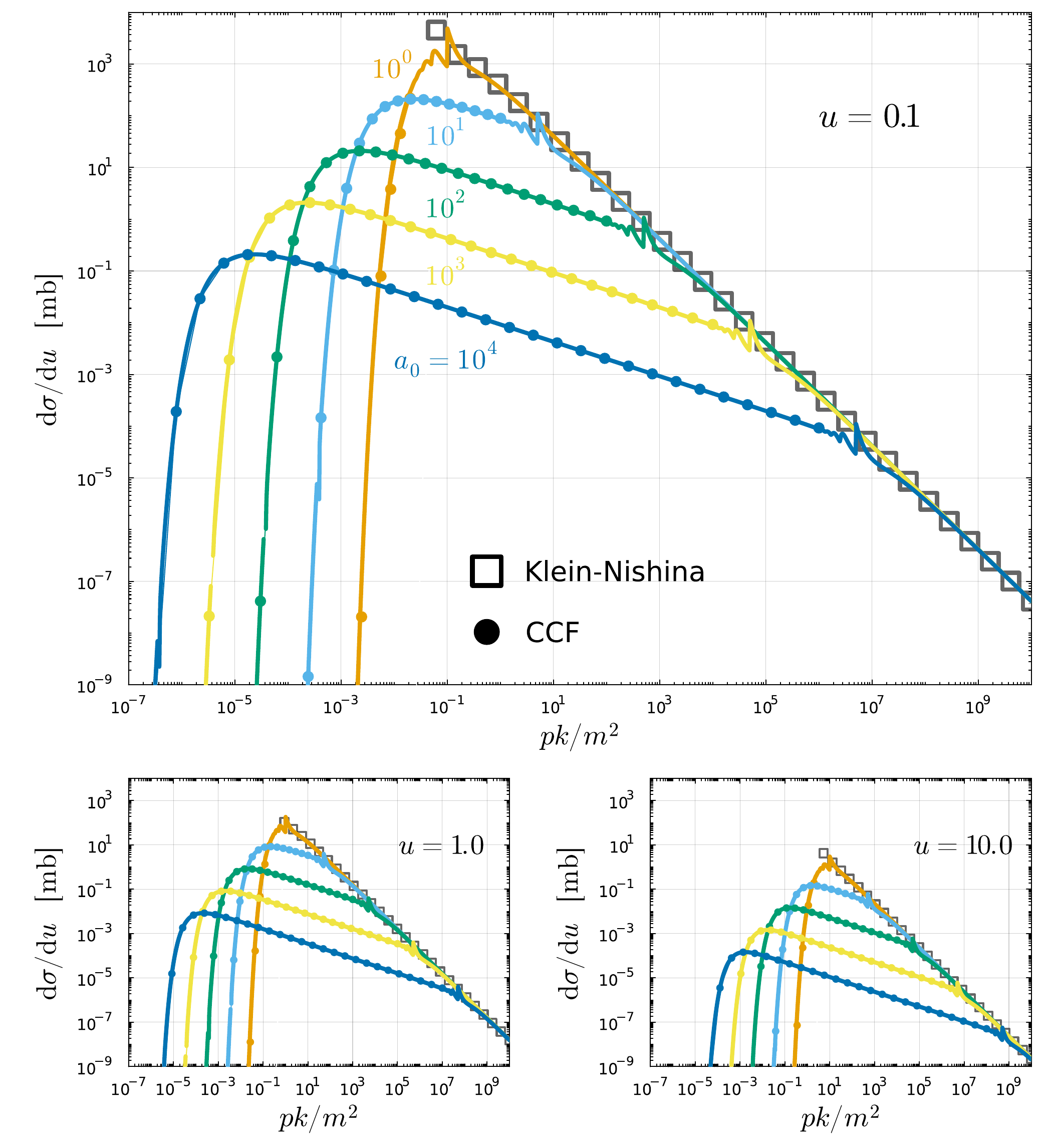}

\caption{Differential IPA cross section (solid curves) for $u = 0.1$ (top panel), 1 (left bottom panel) and 10 (right bottom panel)
as a function of energy for various values of $a_0 = 1, 10^1, 10^2, 10^3$ and $10^4$
(from top to bottom).
The squares are for KN, and dots depict CCF beyond harmonic structures.
\label{fig.pkm2dep}
}
\end{figure}

Turning now to the energy dependence we exhibit in Fig.~\ref{fig.pkm2dep} the differential
IPA cross section for $u = 0.1$ (top panel), $u = 1$ (left bottom panel) and $u = 10$ (right bottom panel).
For a given value of $a_0$, the IPA cross section matches the CCF expression at small energies,
while the high-energy tails coincide with KN. The demarcation is again provided by the first
harmonic structures depending on $a_0$ and $p\cdot k/m^2$ at a given value of $u$ via
$u = u_1 = \frac{2}{1 + a_0^2} p\cdot k/m^2$.
As mentioned above, besides the total cross section, the differential KN cross section also drops
with increasing energy $p \cdot k/m^2$. On top of that dropping is the suppression of the differential
IPA and CCF cross sections at small energies.  

As an interim summary, we mention:
$d \sigma^\mathrm{IPA} / d u \stackrel{\mathrm{small}~p\cdot k/m^2}{ \xrightarrow{\hspace*{1.5cm}}}  
d \sigma^\mathrm{CCF} / d u$,
$d \sigma^\mathrm{IPA} / d u \stackrel{\mathrm{large}~p\cdot k/m^2}{ \xrightarrow{\hspace*{1.5cm}}}  
d \sigma^\mathrm{KN} / d u$
offer a different perspective when keeping fixed $u$ and $a_0$.
Here, ``small energies" mean $p \cdot k/m^2 < \frac12 u (1+a_0^2)$,
while ``large energies" mean $p \cdot k/m^2 > \frac12 u (1+a_0^2)$.
Since this is the edge of the first harmonic, $p\cdot k/m^2$ needs to be 
somewhat smaller for CCF and somewhat bigger for KN.

\section{The $\mathbf{a_0}$ dependence}

\begin{figure}[h!]
\centering
\includegraphics[width=0.7\linewidth]{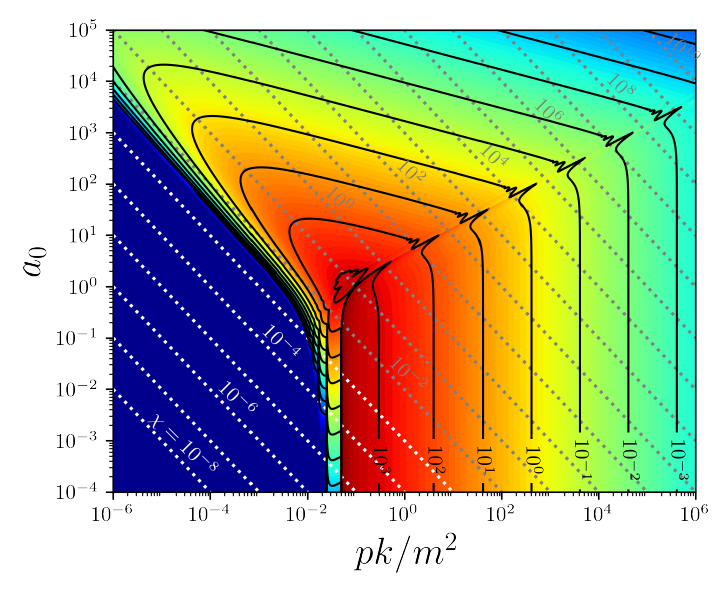} \\ 
\includegraphics[width=0.39\linewidth]{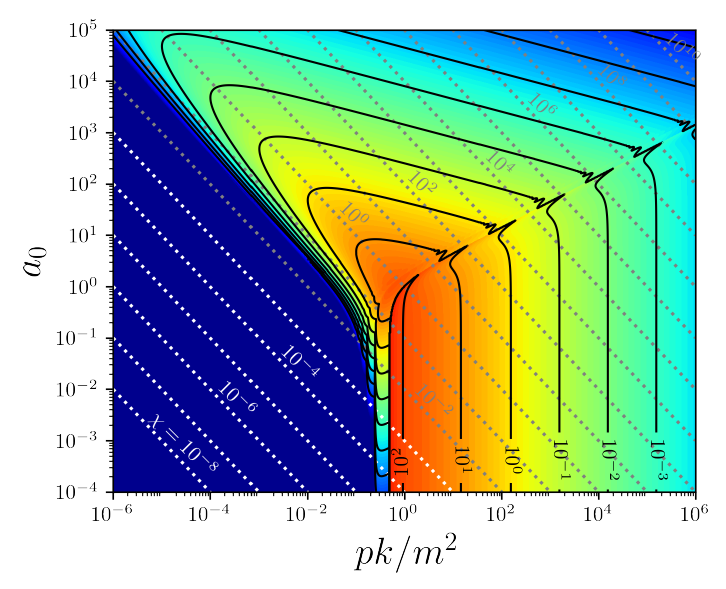} 
\includegraphics[width=0.39\linewidth]{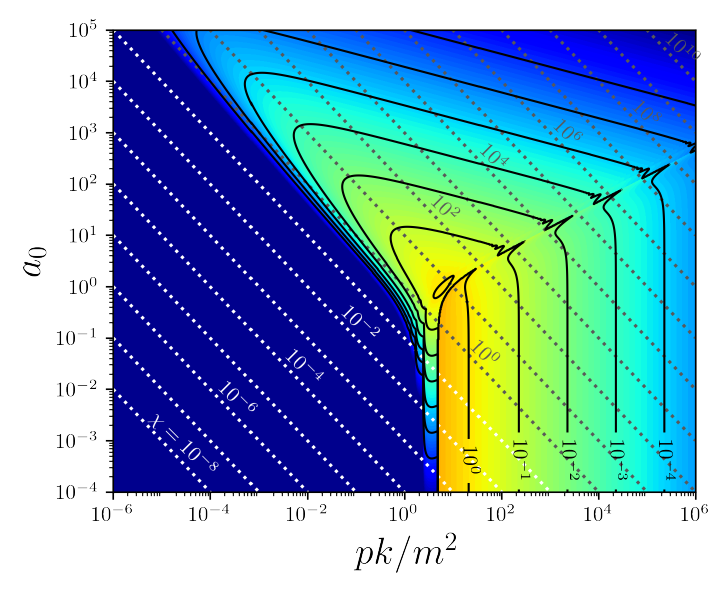} \\
\caption{Contour plots  of $d \sigma^\mathrm{IPA} / du$ in units of mb
over the $a_0$-$p\cdot k /m^2$ pane. Lines of $\chi = const$ are also displayed (dotted).
Top/left bottom/right bottom panels are for $u = 0.1/1/10$.
\label{fig.contour}
}
\end{figure}

To identify further regions of dominating CCF behavior and discriminate against the KN regime,
we display in Fig.~\ref{fig.contour} 
the differential cross section at $u = 0.1$ (top panel), $u = 1$ (left bottom panel) and $u = 10$ (right bottom panel)
as contour plots over the $a_0$-$p\cdot k/m^2$ plane.
The differential cross section clearly shows that even for such large values of $\chi$ as $10^4$
one encounters the KN regime at large energy $p\cdot k/m^2 > 10^2$. The KN regime is that
where the contour lines are vertical, i.e.\ independent of $a_0$.
In contrast, even for small values of $\chi$, e.g.\ $10^0$, the inclined contour lines at small energies,
e.g.\ $p\cdot k/m^2 = 10^{-1}$, point to the CCF regime.
The line $u_1 = u$, at given $u$, is the demarcation, with a structure related to the lowest harmonics. (Appendix \ref{app.D} focuses on the harmonic structures.)\\
Left to the dashed vertical line the KN regime terminates due to kinematic reasons.  
In the deep blue region, the exponential suppression (\ref{eq.9}) applies.

Going, at $p \cdot k/m^2 = const$, from the KN region to larger values of $a_0$ in the CCF region,
one crosses the inclined contour lines, meaning a dropping $d \sigma^\mathrm{IPA} / du$. 
This, together with the $u$ dependence, is at the origin of the dropping total cross section with
increasing $a_0$ exhibited in Fig.~\ref{fig.1}-left.

Coming back to Figs.~2 and 3, one infers that increasing values of $a_0$ suppress the differential
cross section (see Fig.~2 for $p \cdot k/m^2 = const$ and Fig.~3 for $u = const$) at medium-$u$,
while enabling non-zero contributions beyond the KN limit, i.e.\ at $u > u_\mathrm{KN}$.
The net effect is nevertheless the suppression of the total cross section, as exhibited in Fig.~1-left.
That is, the Klein-Nishina effect is amplified by the external field. Both, differential and total cross section
are modified. Reference \cite{Lotstedt:2008} reports on a different example, where an external field modifies the differential cross section, while the total cross section remains nearly unchanged.  This is similar
to the findings in \cite{Seipt:2013hda}.
The difference is in field-enabled vs.\ field-assisted processes.

As an interim summary let us mention that
a more differential perspective, useful for defining ``KN regime" and ``CCF regime" at $u = const$,
is Fig.~4: The KN regime is below the line $u_1 (a_0, p\cdot k/m^2) = u$,
i.e.\ $a_0^{(1)} = \sqrt{\frac{2}{u} \frac{p \cdot k}{m^2} -1}$,
which becomes a near-vertical line $p \cdot k/m^2 = u/2$ for $a_0 < 1$.  The 
contour lines there are vertical, i.e.\ independent of $a_0$.
The CCF regime, where contour lines are inclined due to the $a_0$ dependence,
is above the curves $a_0^{(N)} = \sqrt{\frac{2 N}{u} \frac{p \cdot k}{m^2} -1}$.
In the boomerang-shaped corridor
$a_0^{(1)}(p \cdot k) \cdots a_0^{(N)}(p \cdot k)$, the first $N$ harmonics can be recognized;
their visible structures fade away for $N > 6$, see also Appendix \ref{app.D}.
The CCF regime is apparently limited to $\chi > u/10$ at smaller values of $p \cdot k /m^2$ and larger $a_0$.
Note the back bending of contour lines at $\chi \approx  u$
which approach $\chi = u/10$ at $a_0 > 1$. 
The color code suggests a Y-shaped structure, with the deep blue region belonging to the
exponential suppression, see also the left flanks in Fig.~3.  
(Since $u$ is prescribed and the available energy $p \cdot k/m^2$ in the entrance channel is
too small, even high harmonics do not noticeably contribute to the differential cross section, so that
the CCF region becomes V-shaped as the upper part of the Y.)   

This supplements the summary of the previous section: 
When keeping fixed $u$ and $p \cdot k/m^2$, then
$d \sigma^\mathrm{IPA} / d u \stackrel{\mathrm{large}~a_0}{  \xrightarrow{\hspace*{1cm}}}  
d \sigma^\mathrm{CCF} / d u$,
$d \sigma^\mathrm{IPA} / d u \stackrel{\mathrm{small}~a_0}{  \xrightarrow{\hspace*{1cm}}}  
d \sigma^\mathrm{KN} / d u$.

The $a_0$-$p \cdot k/m^2$ plane can be divided in three regions,
\setlength{\unitlength}{1mm}
\thicklines
\begin{picture}(15,7)
\put(4.5,-1){{{\sf \huge Y}}}
\put(9,0.5){{\footnotesize KN}}
\put(0,3){{\tiny exp.}}
\put(0,0.5){{\tiny supp.}}
\put(5,5.5){{\tiny CCF}}
\end{picture} 
at $u= const$, all captured by $d \sigma^\mathrm{IPA} /du$ at once. Similarly, this was envisaged in \cite{Khokonov:2005}, but is here both quantified and expressed differentially in \( u \).
Such a survey might help discuss the applicability domains of LCFA and LMA
which however require the $u$ integration to arrive at total rates, as pointed out in footnote \ref{fn.6}.
Our consideration suggests that, on a differential level, LMA turns also into LCFA at high intensities.  

\section{Accuracy analysis for IPA vs CCF}
\begin{figure}[b]
\includegraphics[width = \textwidth]{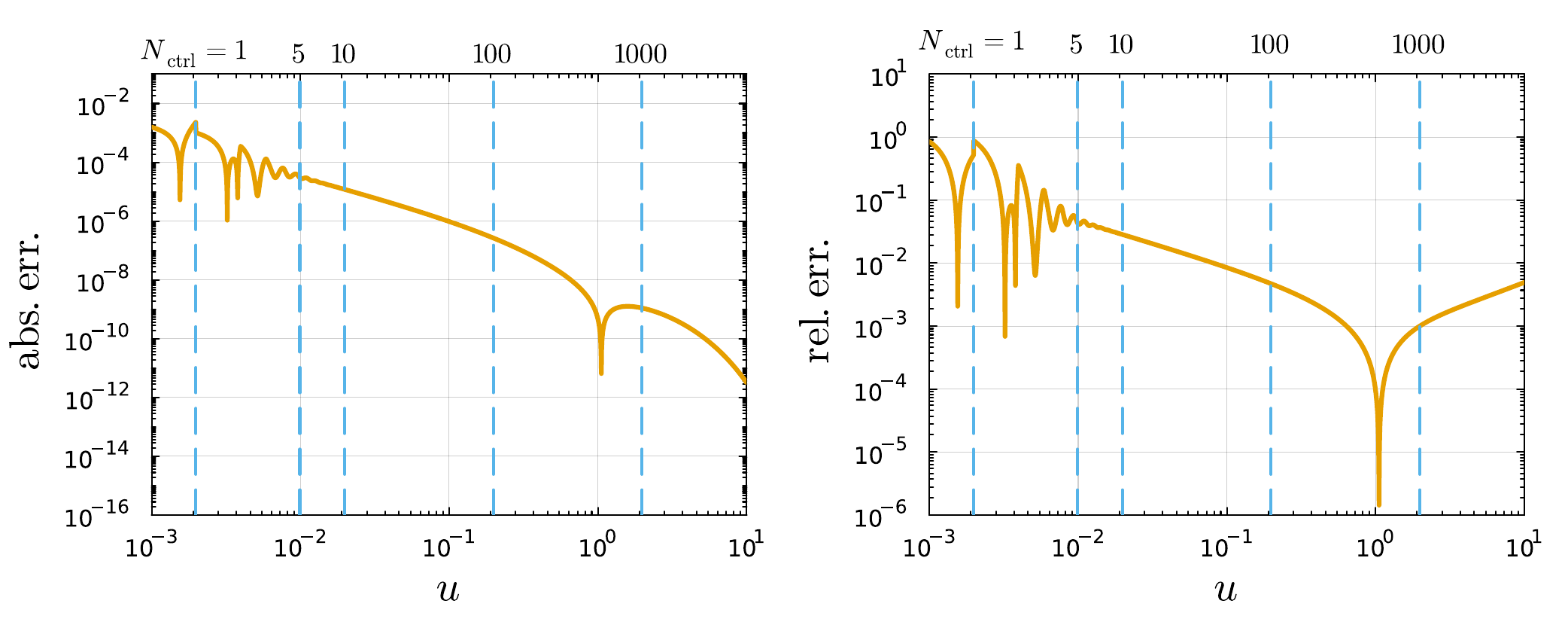}
\caption{Example of absolute (left) and relative error (right) as a function of $u$ for $p \cdot k /m^2 = 0.1 $ and $a_0 = 10$.}
\label{fig:approx_error}
\end{figure}
\begin{figure}[t]
\includegraphics[width = \textwidth]{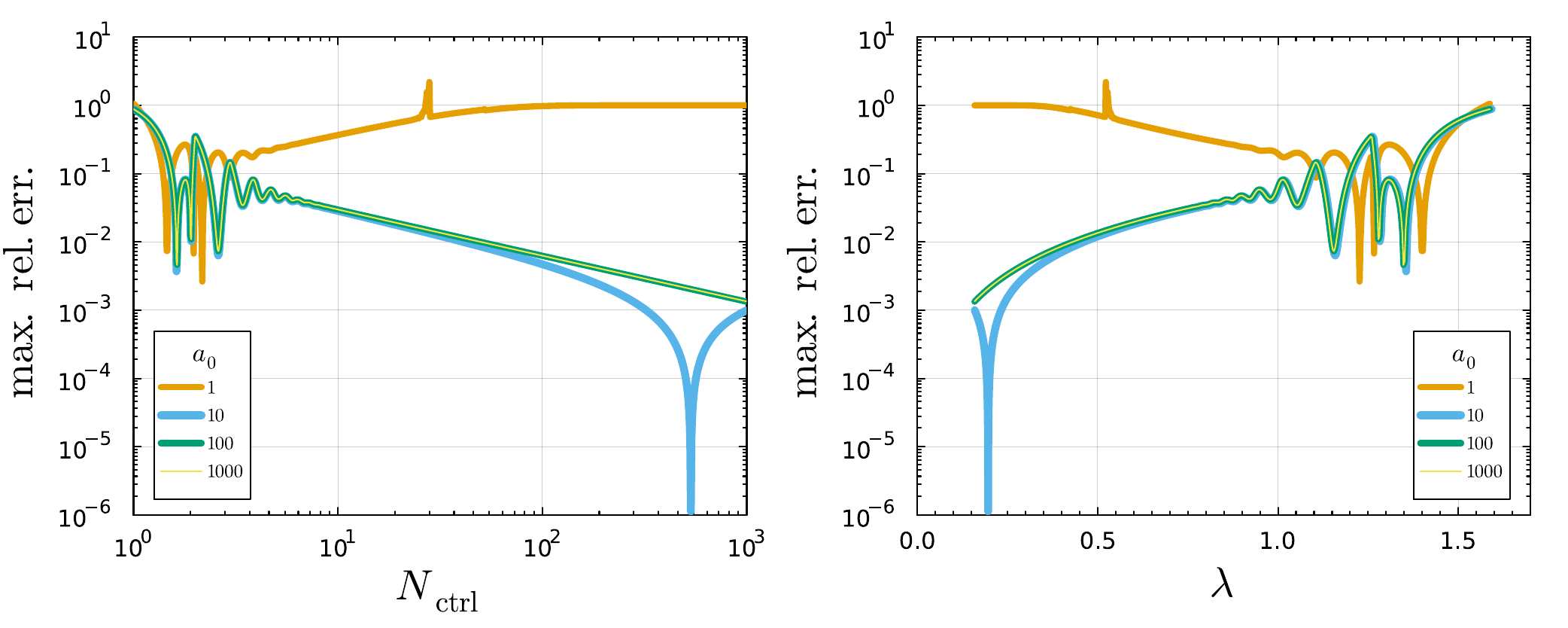}
\caption{ Example of the maximum relative error as a function of $\Nth$ (left) and $\lambda$ (right) for various values of $a_0 = 1, 10, 100, 1000$, and $p \cdot k / m^2  = 0.1$}
\label{fig:max_error}
\end{figure}

Besides being lengthy examined in previous work \cite{DiPiazza:2018bfu, DiPiazza:2017raw, Ilderton:2018nws, Heinzl:2020ynb}, the validity of the replacement  \( \mathrm{IPA} \to \mathrm{CCF} \) for given triple \( (a_0, \chi, u) \) can be controlled for the differential cross-section by the condition
\begin{align}\label{eq:ccf_cond}
\frac{u}{\Nth} \geq \frac{2 \chi}{a_0(1+a_0^2)} = \frac{p \cdot k}{m^2 }\frac{2}{1 + a_0^2},
\end{align}
where \( \Nth \) is the number of harmonics included in IPA (cf.~Eq.~\eqref{eq.5}). 
Technically, the usage of CCF (instead of IPA), \eg in a simulation code, for each triple \((a_0, \chi, u)  \) where \eqref{eq:ccf_cond} is fulfilled, introduces an approximation error controlled by \( \Nth \).\\
 In Fig.~\ref{fig:approx_error}-left, the absolute approximation error \( \mathrm{abs.err.} := \abs{d \sigma^\mathrm{IPA}/d u - \sigma^\mathrm{CCF}/d u } \) is exemplarily exhibited.  
It decreases for \( u \geq \uth = \frac{p \cdot k}{m^2 }\frac{2\Nth}{1 + a_0^2} \).
 This approves again the statement that CCF becomes increasingly better for larger \( u \).  
 Accordingly, in Fig.~\ref{fig:approx_error}-right, this is quantified using the relative approximation error \( \mathrm{rel.err.} = \frac{\mathrm{abs. err.}}{d \sigma^{\mathrm{IPA}}/du} \). For instance, including only \( 10 \) harmonics, i.e., \( N=10 \), in the IPA (see l.h.s.\ of the respective dashed curve) and using CCF on the r.h.s.\ of this dashed curve introduces a maximum error in the single-digit-percent range and improves\footnote{The increasing relative error in Fig.~\ref{fig:approx_error}-right for \( u > 1\) has no physical meaning and results from the imprecise numerical cancelations.} again for larger \( u \).\\
Consequently, for a given \( \Nth > 5 \), the maximum approximation error is given as the value at the respective lower boundary for \( u \), i.e.\ the error value for \( (a_0, \chi, u) \) which solved the equality of \eqref{eq:ccf_cond}. 
Fig.~\ref{fig:max_error}-left shows, that, for \( a_0>1 \), the maximum relative approximation error decreases with increasing  \( \Nth \).
 Furthermore, it is revealed that the error for a given \( \Nth > 5 \) does not change drastically, besides numerical cancelations, for a wide range of \( a_0>1 \), which shows the universality of the control parameter \( \Nth \).
  This can also be observed in the contour plots in Fig.~\ref{fig.contour}, where a straight line gives the lower boundary of the CCF area in the transition to the IPA.
    As already discussed in \cite{Ilderton:2018nws}, in the realm of the local constant-field approximation, the transition from IPA and CCF can be derived analytically by expanding \( \mathrm{d}\sigma^{\mathrm{IPA}}/\mathrm{d} u \) in the parameter \( \lambda := \left(\frac{8 pk}{m^2 a_0^2 u}\right)^{\frac{1}{3}} \) and only keep terms linear in \( \lambda \). Then, the replacement of IPA with CCF is valid\footnote{Similarly, in \cite{DiPiazza:2018bfu,DiPiazza:2017raw}, the parameter \( \etaLCFA :=  \frac{ \chi}{a_0^3 u}\) is proposed to check the validity of LCFA, which is applicable if \( \etaLCFA \ll 1 \). In this regard, using condition \eqref{eq:ccf_cond} leads to \( 1  \gg 1/\Nth \geq 2\chi /(a_0(1+a_0^2)u) \approx \chi/(a_0^3 u) = \etaLCFA \) for \( \Nth >1 \) and large \( a_0 \gg 1 \).}  if \( \lambda\ll 1 \). \\
Using the solution of the equality of the condition \eqref{eq:ccf_cond}, the expansion parameter \( \lambda \) can be numerically related to the maximum relative error. 
In Fig.~\ref{fig:max_error}-right, we observe that even if \( \lambda \) is relatively large, say \( \sim 0.7 \), the introduced maximum error by using CCF instead of IPA is still in the single-digit percent range. 
Therefore, by using the condition \eqref{eq:ccf_cond} and especially the control parameter \( \Nth \), one can quantify, what the "\( \ll\)" in \( \lambda \ll 1 \) actually means.

\section{Many roads lead to large \boldmath{$\chi$} }

The $a_0$-$p \cdot k/m^2$ plane can be mapped out by many other coordinates,
e.g.\ $\chi = a_0 \, p \cdot k/m^2$ plus a second one, such as $a_0$ or $ p \cdot k/m^2$ or
(functions of) combinations thereof. An example of orthogonal coordinates is the pair
$\chi, \eta^2$ with $\eta^2(a_0, p \cdot k/m^2) = (p\cdot k/m^2)^2 - a_0^2$ 
for $\eta^2 \in [-\infty, \infty]$ 
and our $\chi(a_0, p \cdot k/m^2) = a_0 \, p\cdot k/m^2$ for $\chi \in [0, \infty]$.

\begin{figure}[h!]
\includegraphics[width=0.49\linewidth]{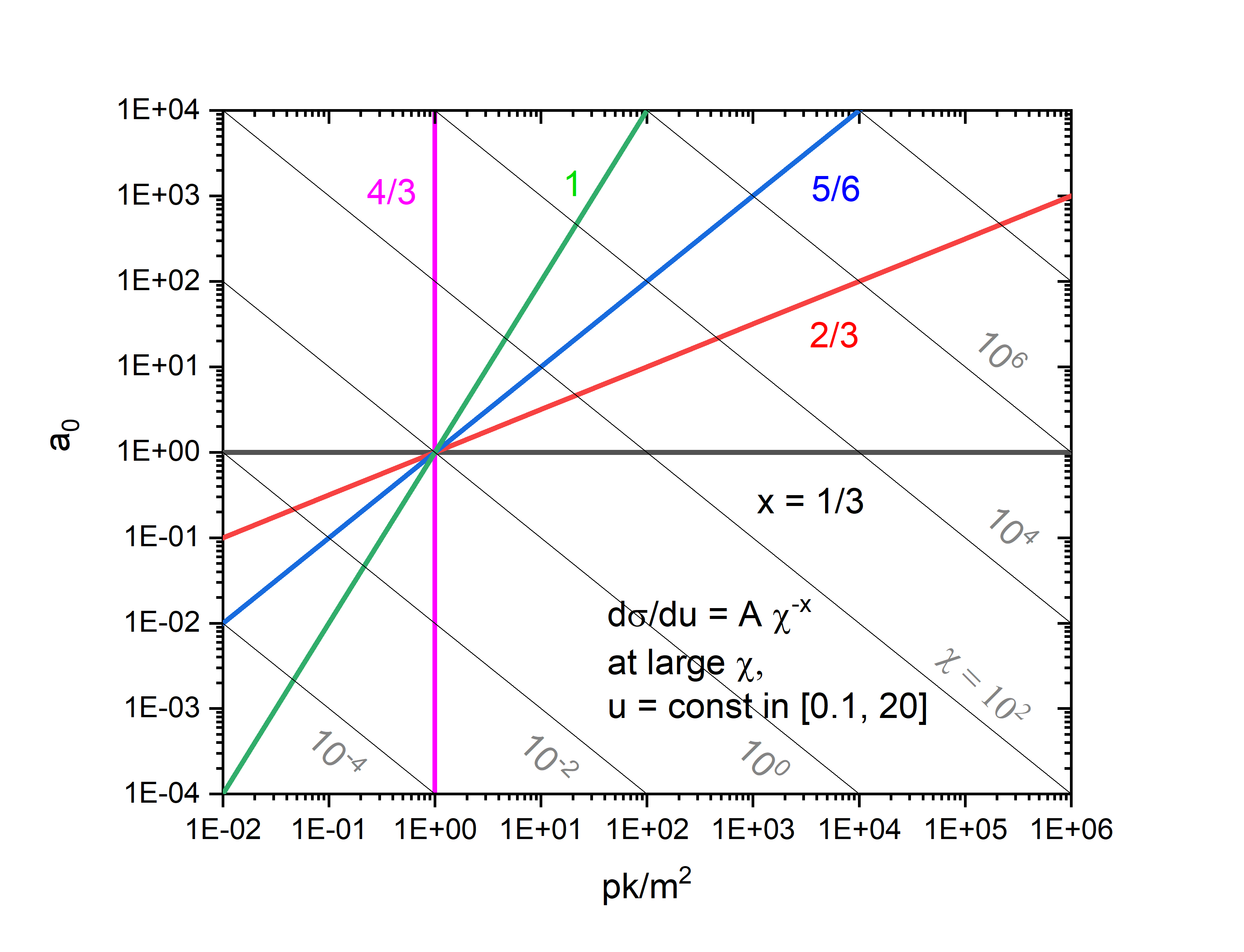}
\includegraphics[width=0.49\linewidth]{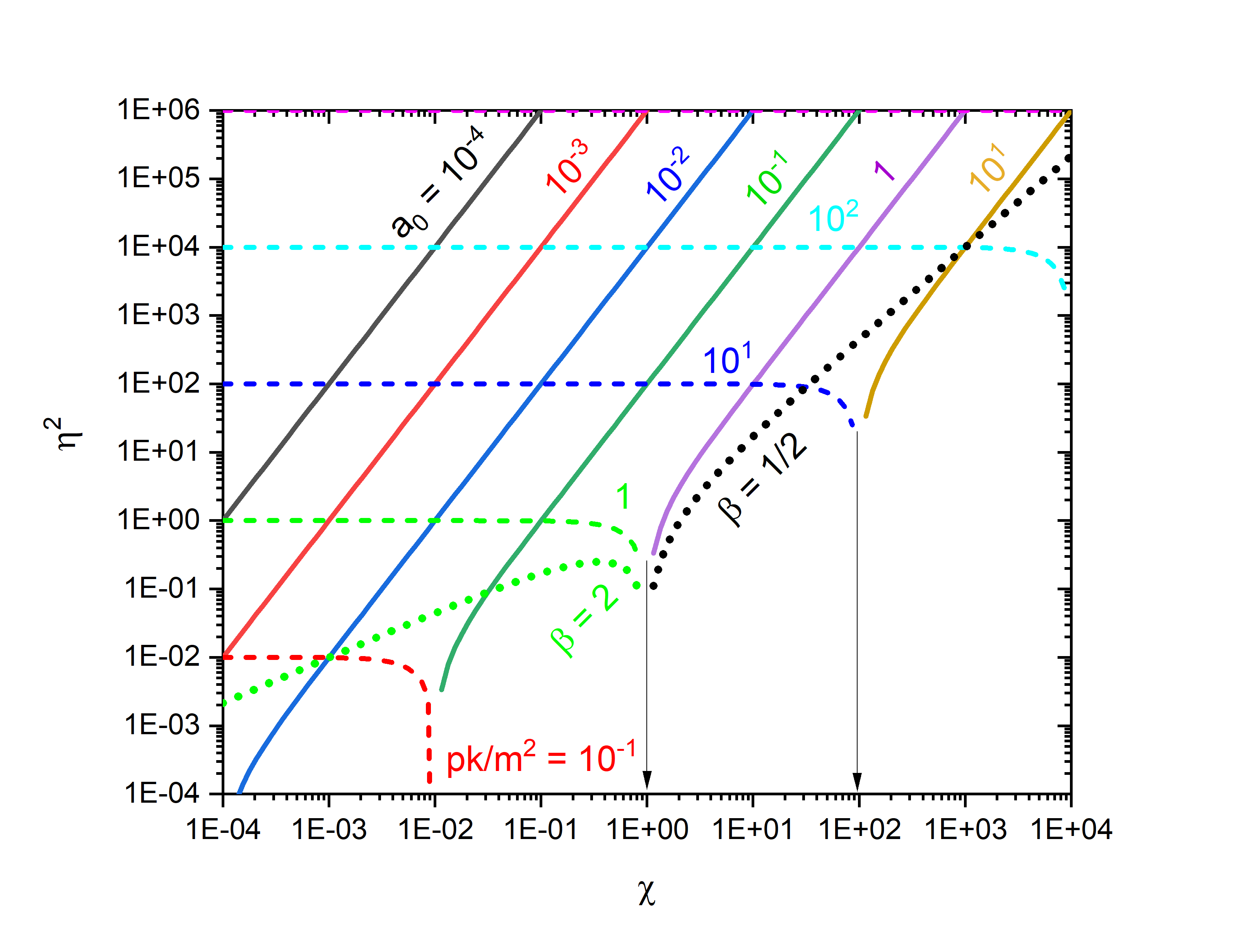}
\put(-143.5,57.5){$\beta = \color{magenta}\infty\color{black}$}
\put(-121.5,57.5){$\color{green}2\color{black}$}
\put(-107.5,57.5){$\color{blue}1\color{black}$}
\put(-91.5,47.5){$\color{red}1/2\color{black}$}
\put(-91,30.5){$\color{black}0\color{black}$}
\caption{Left panel: The $a_0$-$p \cdot k/m^2$ plane is mapped out by curves $\chi = const$ (grey lines).
A few paths of going to large values $\chi$ are depicted, labeled by the parameter $x$
determining the asymptotic  $\frac{d \sigma^\mathrm{CCF}}{du}\vert_{u=const} \propto \chi^{-x}$.
Right panel: The same information but in orthogonal coordinates $\eta^2 > 0$ vs.\ $\chi$.
Solid and dashed curves are for $a_0 = const$ and $p\cdot k/m^2 = const$, respectively.
(The near-vertical sections are replaced by arrows.)
Heavy dots depict the trajectories with $\beta = 1/2$ and 2; the trajectory with $\beta = 1$ is along the
$\eta^2 = 0$ line (not exhibited). Mapping out the full $a_0$ vs.\ $p\cdot k / m^2$ plane displayed in the
left panel requires the complement $\eta^2 < 0$ vs.\ $\chi$ (not shown since it arises from re-labeling 
$a_0 \leftrightarrow p\cdot k/m^2$ for the coordinate lines $a_0 = const$ and $p \cdot k/m^2 = const$
as well as $1/2 \leftrightarrow 2$ for the trajectories).
\label{fig.plane}
}
\end{figure}

Trading CCF as proxy of $\frac{d \sigma^{\mathrm{IPA}}}{du}$,
the differential cross section can be fitted by 
$\frac{d \sigma^\mathrm{CCF}}{du}\vert_{u=const} = \mathcal{A}(u) \, \chi^{-x}$
for $u \in [0.1, 20]$ in the asymptotic region, i.e.\ at large $\chi$.
Various trajectories of going to large $\chi$
in the $a_0$ vs.\ $p\cdot k/m^2$ plane are exhibited in Fig.~\ref{fig.plane}-left,
for definiteness starting at $a_0 =1$ and $p \cdot k/m^2 = 1$.
We find for a path along $a_0 = const$ (black line): $x = 1/3$,
along $a_0 = (p\cdot k/m^2)^{\beta}$: $x = 2/3$ for $\beta = 1/2$ (red line),
$x = 5/6$  for $\beta =1$ (blue line), 
$x = 1$ for $\beta = 2$ (green line), and
along $p \cdot k/m^2 = const$ (magenta line): $x =4/3$. 
That is, $x$ depends strongly on the path of going to large $\chi$. 
For a path $a_0 = \mathfrak{z} \left( \frac{p\cdot k}{m^2} \right)^\beta$, one gets the $\chi$ dependence
$ \frac{d \sigma^\mathrm{CCF}}{du} \propto \mathfrak{z}^{- 1/(1+\beta)}\chi^{-1/3 - \beta/(1+\beta)}$.
The large-$\chi$ limits considered in  \cite{Ilderton:2019kqp,Podszus:2018hnz}
are contained as special cases.

Figure \ref{fig.plane}-right exhibits the same information in orthogonal coordinates $\eta^2 > 0$ vs.\ $\chi$.
For $p\cdot k/m^2 > a_0$, $\eta^2 \approx const > 0$ corresponds to going to larger $\chi$
at $p\cdot k/m^2 = const$, and {\it vice versa} $\eta^2 \approx const < 0$ corresponds to
going to larger $\chi$ at $a_0 = const$ as long as $a_0 > p\cdot k /m^2$.
(See remark in figure caption on the relation of $\eta^2  \lessgtr 0$ sectors.)
As said above, many ways lead to large $\chi$ (and result in fairly different asymptotic regimes).

In a nutshell: along trajectories in the $a_0$-$p\cdot k/m^2$ plane, 
we find $d \sigma^\mathrm{CCF} \propto \chi^{-x}$
with $x$ strongly depending on the actually chosen path in traversing the lines $\chi = const$.
In that region, where CCF is a good approximation of IPA, this finding applies equally well to
$d \sigma^\mathrm{IPA} / du \vert_{u = const}$.

\section{Discussion of dependencies and regimes}\label{sect.VII}

The here employed formulas (\ref{eq.1} - \ref{eq.12}) of one-photon emission in some external 
univariate null-fields
display often structures such as $1 + u$, $1 + a_0^2$, $1 - u/u_n$,  or $1 + u_\mathrm{KN}$ etc.\
suggesting various regimes depending on whether $1 < u$ or $1 > u$,  $1 < a_0^2$ or $1 > a_0^2$,
$u_n < 1$ or $u_n > 1$ etc. 
Our numerical study has the goal of identifying the corresponding regimes, beyond asymptotic behaviors,
such as $u \to 0$, $a_0 \to 0$ or $a_0 \to \infty$, $p\cdot k/m^2 \to \infty$ etc.
That is, we quantify the triple $u$ - $a_0$ - $p\cdot k/m^2$ for moving in upper part of the triangle
\setlength{\unitlength}{1mm}
\thicklines
\begin{picture}(24,12)
\put(-0.5,-1){$d \sigma^\mathrm{KN}$} 
\put(13,-1){$d \sigma^\mathrm{CCF}$} 
\put(6,9){$d \sigma^\mathrm{IPA}$} 
\put(8.0,8){\vector(-1,-1){5}}
\put(8.0,8){\vector(1,-1){5}}
\put(4,-1){\line(1,0){8}}
\end{picture}
with baseline $d \sigma^\mathrm{KN} \leftrightarrow d \sigma^\mathrm{CCF}$ 
given by classical emission (e.g.\ described by photon emission along an electron trajectory in the 
external field via Lienard-Wiechert potential which ignores the recoil).
A key for comparing IPA and KN and CCF is the proper definition of their differential cross sections,
including a suitable definition of the intensity parameter $a_0$ in CCF.
Roughly speaking, CCF applies at $F \frac{2}{u} \frac{1}{1 + a_0^2} \frac{p \cdot k}{m^2} <1$ and
KN at $F^{-1} \frac{2}{u} \frac{1}{1 + a_0^2} \frac{p \cdot k}{m^2} >1$, where $F = 1 + \mathcal{O}(1)$ 
is a fudge factor aimed at accommodating the region of first harmonics.
In other words, the three-dimensional parameter space has a layer from
$u = F^{-1} \,  \frac{2}{u} \frac{1}{1 + a_0^2} \frac{p \cdot k}{m^2}$ to 
$u = F \, \frac{2}{u} \frac{1}{1 + a_0^2} \frac{p \cdot k}{m^2}$; above (below) that layer, 
CCF (KN) approximates fairly well the invariant differential IPA cross section.

Although, it can be analytically proven \cite{Ritus85}, it is somewhat surprising that 
$d \sigma^\mathrm{IPA}_\mathrm{circ.\, pol.} \to d \sigma^\mathrm{CCF}$, since
circular polarization (circ.\ pol.)
means a rotating electric field, i.e.\ one would expect a closer correspondence 
of $d \sigma^\mathrm{IPA}_\mathrm{lin.\, pol.}$ and $d \sigma^\mathrm{CCF}$
since CCF means fixed (space-time independent) electric ($E$) and magnetic ($B$) fields,
which can be aligned with the linearly polarized (lin.\ pol.) wave fields.
In Appendix \ref{app.B}, this issue is examined, along with a discussion of the meaning of $a_0$ w.r.t.\ CCF.
 
We prefer to consider the differential probability in the invariant Ritus variable $u$ instead of
using the scattering angle or frequency $\omega'$ of the emitted photon. 
For the mapping, $u \mapsto \omega'$, cf.\ the discussion in \cite{Acosta:2021iyu} and Appendix \ref{app.C}.

When turning from the monochromatic external field to plane-wave \underline{pulses}, the differential spectra become
smoothed out, as quantified in \cite{Kampfer:2020cbx}, i.e.\ the harmonic structures are less significant. 
In momentum balance equations, the electron IPA quasi-momenta are replaced by momenta, 
as occurring in CCF and KN, of course. 
 
\section{Summary}

Inspired by the ongoing discussion of the Ritus-Narozhny conjecture concerning the validity
of the Furry picture for processes at and in external fields, we analyze numerical probabilities
of linear and nonlinear Compton scattering. To establish a link to the Klein-Nishina cross section
for linear Compton scattering, we chose suitably normalized probabilities
as ``cross sections", mainly based on the
long-known relation of the weak-field limit of the nonlinear Compton scattering matching the Klein-Nishina
expression. Choosing furthermore the monochromatic plane wave model as a proxy of a mono-energetic
photon beam, we show with established expressions\footnote{Since we envisage a clear-cut picture with well-defined precise basics, we resort to the simplest (quantum) formulas describing exclusive one-photon emission. Thus, such a compendium ignores much of the refinements elaborated during the last years, e.g. w.r.t. pulse shapes (implying bandwidth, ponderomotive and interference effects), laser focusing, and the many useful approximation schemes and simulation codes, etc., including the efforts of benchmarks. By exhibiting a few relevant plots, we want to display systematically in an intriguing manner the interplay of Klein-Nishina (i.e. linear Compton) and constant-crossed fields to nonlinear Compton and identify - purely numerically - the overlap regions. For a comprehensive survey of modifications of the IPA model, see \cite{Fedotov:2022ely}. In essence, the parameter space is noticeably enlarged requiring a multitude of specialized considerations of various distinctions.} for the one-photon emission in pane waves
and constant crossed field how and where the so-defined cross sections coincide.
We do our analysis differentially with a Lorentz invariant quantity - the Ritus variable $u$ - 
instead of frequency or/and scattering angle of the produced photon 
emitted by an electron moving in the external field or scattering at a photon beam.
We identify those regions in the three-dimensional parameter space 
$u$ (for the emitted photon) --  $a_0$ (external field intensity) -- $p\cdot k/m^2$ (energy),
where the constant crossed field approach is trustable. Since the differential cross section matches
inevitably the Klein-Nishina cross section at certain (small) values of $u$, the total cross section contains
also such a contribution. We argue that in discussing the Ritus-Narozhny conjecture one should
resort to such a differential consideration. 
It could be that the tails of the Klein-Nishina behavior require adequate treatment in simulation codes,
far beyond the improved constant cross field approximations.

For significant external field intensities, $a_0 > 1$, the Klein-Nishina effect, i.e.\ the suppression of (differential) cross section at high energies, becomes noticeably stronger.   

Among the directions of further improvements are extensions accounting for polarization effects,
e.g.\ along the path paved in \cite{Chen:2022dgo} with various applications
\cite{Seipt:2023bcw,Gong:2022qis,Song:2024oak,Titov:2024aez}.
Much more compatibility (both computationally and w.r.t.\ parameter settings)
is to be faced for nonlinear Compton scattering \underline{within} a background field,
which we envisage in subsequent work.

\begin{appendix}

\section{A few Numbers }\label{app.A}

To justify the use of the considered fairly wide spread of energies and field strengths we quote a few numbers
relevant to high-intensity laser physics and astrophysics.\\
\begin{center}
\begin{tabularx}{\textwidth}{
ll
|c|c|c|c|c|c
| c
 | >{\centering\arraybackslash}X 
}
\toprule
                & & $\lambda\ [\mathrm{nm}]$ & $\omega\ [\mathrm{eV}]$ & $I\ [\frac{\mathrm{W}}{\mathrm{cm}^2}]$ & $a_0$\footnote{
Using $a_0 \approx 60 \frac{\lambda}{\mu\mathrm{m}} \sqrt{I /(10^{22} \mathrm{W/cm}{}^2)}$ with laser intensity $I$ and wave length $ \lambda$.
} & $\ E\ [\mathrm{GeV}]\ $        & $\ pk/m^2\ $ & $\ \ \chi\ \ $ & ref. \\\hline
LUXE &           & 800       & 1.55     & $2\times 10^{21}$     & 21.5    & 17.5       &   0.2077   & 4.5     &       \cite{Abramowicz:2021zja}     \\
\mbox{FACET-II}& &&&&&&&&\\
&\hspace*{-1.5cm} E320 & 800       & 1.55     & $2\times 10^{20}$     & 6.8    & 13         &    0.1543      &   1.05     &      \cite{Meuren:2020nbw}      \\
HED-HIBEF\footnote{The primary aim is at vacuum birefringence investigations by two-laser interactions \cite{Ahmadiniaz:2024xob} with
(i) XFEL: $\omega = 1 \cdots 15$~keV, $a_0 < \mathcal{O}(10^{-3})$ and
(ii) optical laser:  $\omega = 1.55$~eV, $a_0 < 30$. These may optionally be used in collisions with
(a) electron beam $E = 100$~MeV (projection) or 
(b) rest gas electrons $E = \mathcal{O}(1~\mathrm{MeV})$ 
which are useful for calibration purposes in stand-alone configurations
uncovering a wide parameter range, pointing either to the onset of the KN suppression regime, e.g.\ in  (i.a), 
or to the suppression by the laser field in the Thomson regime, e.g.\ in (ii,a).} &           &           &          &                      &       &            &          &        &           \\
&\hspace*{-1.5cm} EuXFEL            & $0.1$       & $13\times 10^3$       & $10^{21}$            & $0.002$ & $0.1$ &    $9.5$      &   $0.02$     &     \cite{EuXFEL}      \\
&            &        &         &             &  & $0.001$ &    $0.09$      &   $1.6\times 10^{-4}$     &           \\
&\hspace*{-1.5cm} RELAX           & $795$       & $1.56$     & $8\times 10^{21}$ & $42.7$ & $ 0.1$ &  0.001        &    $0.05$    &     \cite{RELAX}      \\
&& & & & & $0.001$ &  $1.1\times 10^{-5}$ &  $4.7\times 10^{-4}$ &     \\
ELI-NP  &        & $814$       & $1.52$     & $10^{22}$            & $49.8$    &            &          &        &    \cite{Gales:2018}       \\
CoReLS\footnote{Current maximum for optical lasers.}  &        & $800$       & $1.55$     & $1.1\times 10^{23}$   & $159.2$   &            &          &        &\cite{Yoon:2021ony} \\ \hline\hline
cosmic rays && &$10^6$ &&&$10^4$ & $7.7\times 10^7$ &&\cite{Boncioli:2023gbl}\\
vela pulsar && &$10^{13}$ &&&$ 2\times 10^{4}$ & $1.5\times 10^{15}$ && \cite{HESS:2023sxo} \\
CMB  && &$2.34 \times 10^{-4}$ &&&$ 10^{4}$ & $ 0.0179 $ &&  \cite{ParticleDataGroup:2022pth}
\end{tabularx}%
\end{center}
In general, magnetars produce magnetic fields\footnote{
Critical field strengths: $B_{c} =m^2 c^3 /\hbar e  \approx 4.4 \times 10^{13}$~G
and $E_c = m^2 c^3 / \hbar e \approx  1.323 \times 10^{18}$~V/m.  } with $B \approx \mathcal{O}(10^{14} \cdots 10^{15}~\mathrm{G})$ \cite{Li:2024oda}. In addition to the listed laser facilities, we mention NSF ZEUS \cite{ZEUS}, NSF OPAL \cite{OPAL}, and SEL \cite{SELabc}. For further details, the interested reader is referred to the cited references.

\section{Linear vs. circular polarization and CCF}\label{app.B}

One must not consider CCF as $\omega \to 0$ limit \cite{Ritus85}, but
compare the plane-wave field and CCF at the same field strengths
$E := \sqrt{E_x^2 + E_y^2 + E_z^2}$ and $B := \sqrt{B_x^2 + B_y^2 + B_z^2}$.
Following the reasoning and notation in \cite{Heinzl:2008rh} we recap the definition of a pane wave
by the field strength tensor 
$F^{\mu \nu} (x) = F_1(\omega n \cdot x) f_1^{\mu \nu} +  F_2(\omega n \cdot x) f_2^{\mu \nu}$ with
$f_i ^{\mu \nu}:= n^\mu \epsilon_i^\nu - n^\nu \epsilon_i^\mu$ and
$n^2 = 0$, $n \cdot \epsilon_i = 0$, $\epsilon_1 \cdot \epsilon_2 = 0$,
yielding transversality $n^\mu F_{\mu \nu} = 0$ and 
energy-momentum tensor $T_{\mu \nu} = (\sum_{i=1,2} F_i^2) n_\mu n_\nu$.
The normalized gauge and Lorentz invariant field intensity parameter reads 
$a_0^2 = \frac{e^2}{m^2} \frac{\langle \langle  p_\mu T^{\mu \nu} p_\nu \rangle \rangle}{(p \cdot k)^2}$
with $\langle \langle  \cdots \rangle \rangle$ meaning Lorentz invariant proper-time ($\tau)$ average.
This facilitates the following relations:
\begin{center}
\begin{tabular}{r | c | c | c}
\toprule
   & lin.\ pol. & circ.\ pol. & CCF \\
\hline
$F_1 =  \mathcal C \omega ~\times$ & $\cos k\cdot x $ &  $   \cos k\cdot x $ & 1\\
$F_2 =  \mathcal C \omega ~\times$ & $0 $                  &  $ - \sin k\cdot x $ & 0\\
\hline
$E_x = B_y = \mathcal{C} \omega ~\times$ & $\cos \Omega \tau $ & $\cos \Omega \tau$ & 1 \\
$E_y = B_x = \mathcal{C} \omega ~\times$ & $ 0$ & $\sin \Omega \tau $ & 0 \\
$E = B = \mathcal{C} \omega ~\times $ & $\cos \Omega \tau$ & 1 & 1\\
$T_{\mu \nu} = \mathcal{C}^2 k_\mu k_\nu ~\times$ & $\cos^2 k\cdot x $ & 1 & 1\\
$a_0^2 = \frac{e^2}{m^2} \mathcal{C}^2~\times$ & $\frac12$ & 1 & 1\\
\hline
$b_1 = \mathcal{C} ~\times$ & $1 - \cos \Omega \tau$ & $1 - \cos \Omega \tau$ & $\frac 12 \Omega^2 \tau^2 $\\
        & $\approx \frac12 \Omega^2 \tau^2 (1 + \frac{1}{12} \Omega^2 \tau^2)$ & 
           $\approx \frac12 \Omega^2 \tau^2 (1 + \frac{1}{12} \Omega^2 \tau^2) $ &       \\
$b_2 = \mathcal{C} ~\times$ & $0$                             & $-(\Omega \tau - \sin \Omega \tau) $ & 0 \\
        &   & $\approx - \frac16 \Omega^3 \tau^3 (1 - \frac{1}{20} \Omega^2 \tau^2)$ & \\
$b_3 = \mathcal{C}^2 ~\times$ & $\frac14 (2 \Omega \tau - \sin 2 \Omega \tau) $ & 
$ 2  (\Omega \tau - \sin \Omega \tau)$ &  $\frac13 \Omega^3 \tau^3 $\\
   & $\approx \frac13 \Omega^3 \tau^3 (1 - \frac{1}{5} \Omega^2 \tau^2)$ & 
$\approx \frac13 \Omega^3 \tau^3 (1 - \frac{1}{20} \Omega^2 \tau^2) $ & \\
\end{tabular} 
\end{center}
where $\Omega \tau := k \cdot x$ and $k = \omega n$. 
(We use coordinates with $n = (1, 0, 0, 1)$, $\epsilon_1 = (0, 1, 0, 0)$, $\epsilon_2 = (0, 0, 1, 0)$,
hence $E_z = B_z = 0$.)
The equality of $T_{\mu \nu}$, $E$ and $B$, and $a_0^2$ for circular polarization and CCF
supports their close correspondence.
The electron trajectories determine the classical emission. They are described by
$x^\mu(\tau) = x^\mu(0) + p^\mu(0) \, \frac{\tau}{m} - \epsilon^{\mu \, i} b_i(\tau) \frac{e}{p \cdot k}
+ k^\mu b_3 (\tau) \frac{e^2}{2 (p \cdot k)^2}$ with coefficients $b_{1, 2, 3}$ listed above.
In leading order in powers of $\Omega \tau$, all $b_{1, 3}$ agree and $b_2^{\mathrm{circ.~pol.}}$ 
is nonzero but suppressed,
$b_2^{\mathrm{circ.~pol.}} \approx - \frac13 b_1^{\mathrm{circ.~pol.}} \Omega \tau (1 + \frac{1}{30} \Omega^2 \tau^2)$.
The periodicity of the IPA trajectories (figure-8/ellipse in $x$-$z$ plane and a frame, where drift terms vanish,
for lin./circ.\ polarizations) is at the origin of the IPA quasi-momenta, see Appendix in \cite{Seipt:2012tn} for instance.
Note the cancellation of $p\cdot k$ in the definition of $a_0^2$ and 
$a_0 = \frac{e}{m} \mathcal{C} = \frac{m}{\omega} \frac{E}{E_c}$ due to
$\mathcal{C} = E / \omega$ for both circular polarization and CCF.
As emphasized above, $E$ and $\omega$ are primarily given by the considered plane wave and used
afterward to define the related $a_0$ for CCF. Alternatively, one simply follows \cite{Ilderton:2018nws}
and defines local $\chi(\phi)$ and $a_0(\phi)$, thus considering CCF expressions as some approximation
of IPA without recourse to real CCF.

\section{Master curves of harmonic structures}\label{app.D}

\begin{figure}[h!]
\includegraphics[width=0.49\linewidth]{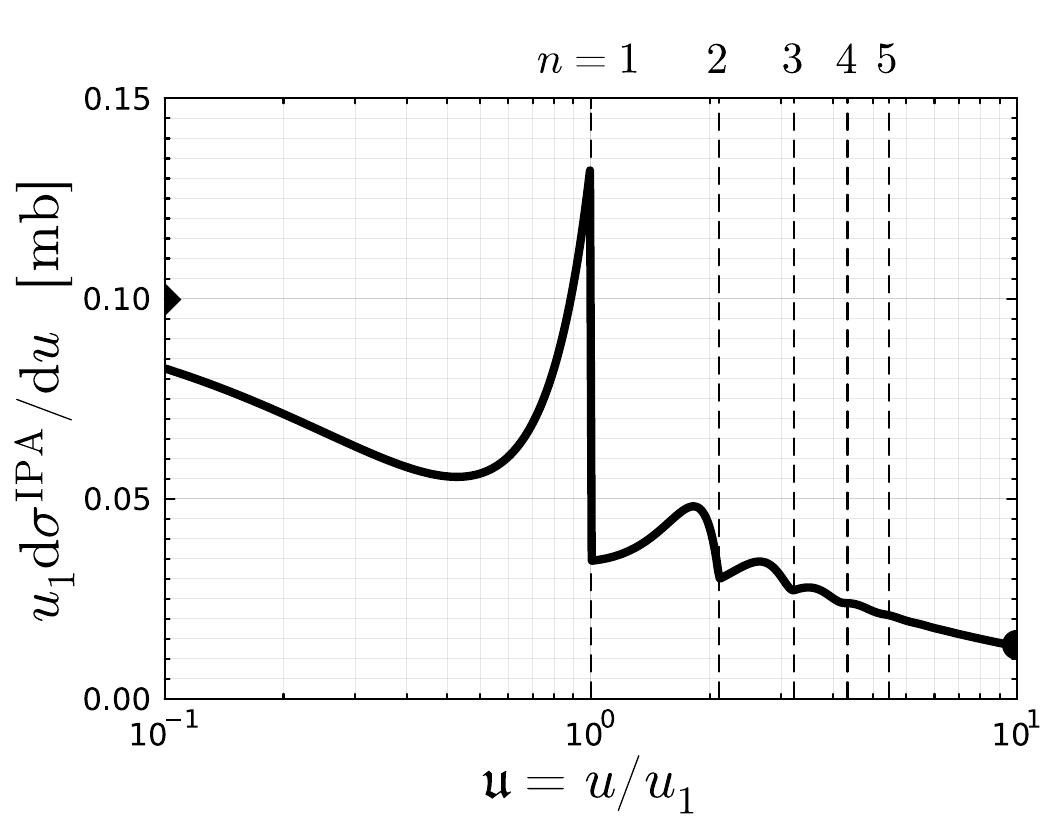}
\includegraphics[width=0.49\linewidth]{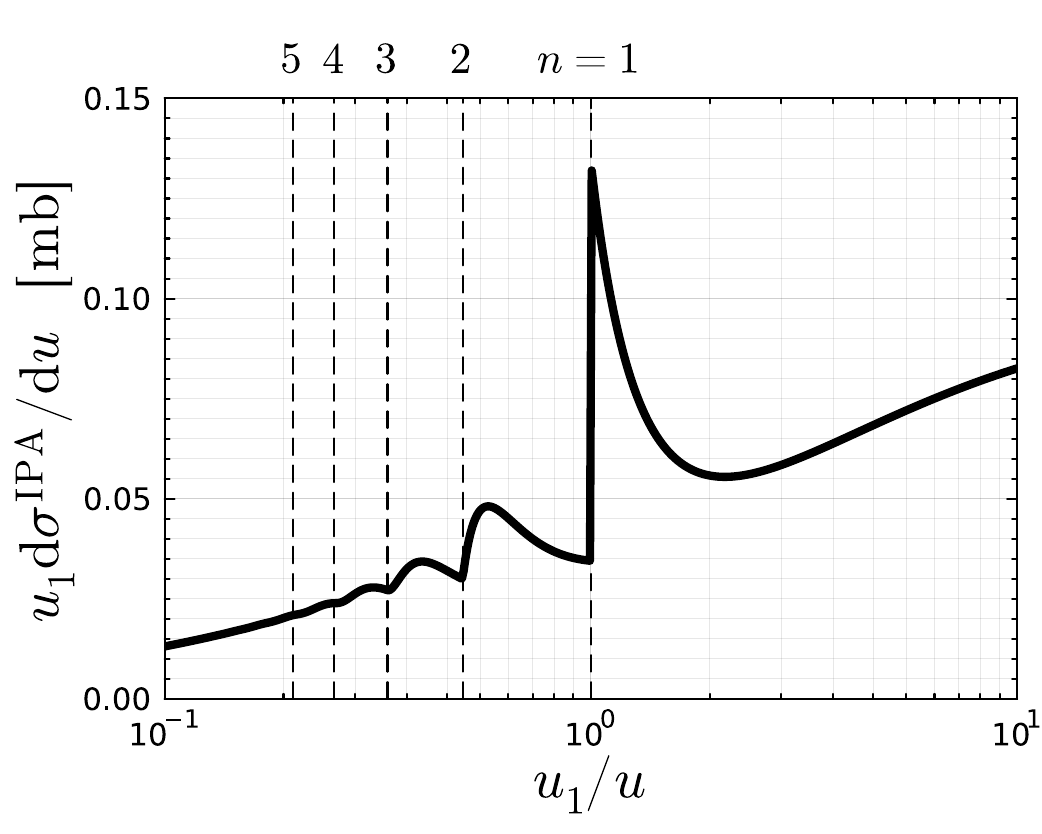}
\caption{Master curves $u_1 \frac{d \sigma^\mathrm{IPA}}{d u}$
displaying harmonic structures on a severe background at $u \ll 1$ and $a_0 \gg 1$
as a function of $u / u_1 = \mathfrak{u}$ (left panel, complementing Fig.~\ref{fig.a0dep})
and $u_1/u=\frac{p \cdot k}{m^2} \frac{2}{u (1+a_0^2)} = \mathfrak{u}^{-1}$ (right panel, complementing Fig.~\ref{fig.pkm2dep}).
One can identify five harmonics $n = 1, \cdots, 5$, with the last one ($n = 5$) being very shallow.  
Due to independent evaluations
(the left panel is for running $u$ with $p \cdot k/m^2 = 10$ and $a_0 = 10^2$,
while the right panel is for running $p \cdot k/m^2$ with $u=10^{-2}$ and $a_0 = 10^2$),
the curves are not perfectly mirror-symmetric, as they should be
in the asymptotic region. 
The l.h.s.\ triangle is for  $\frac{p \cdot k}{m^2} \frac{d \sigma^\mathrm{KN}}{d u}\vert_{u \to 0}$,
and the r.h.s.\ bullet is at $\frac{p \cdot k}{m^2} \frac{d \sigma^\mathrm{CCF}}{d u}\vert_{u / u_1 = 10}$,
both in the left panel only.
\label{fig.M}
}
\end{figure}

For $u \ll 1$ and $a_0 \gg1$, the leading order of $\frac{p \cdot k}{m^2} \frac{d \sigma^\mathrm{IPA}}{d u}$
depends only on $\mathfrak{u} := u/u_1 \approx u a_0^2 m^2 / 2 p \cdot k$
since $z_n \approx 2 \sqrt{\mathfrak{u} (n - \mathfrak{u})} + \mathcal{O}(a_0^{-2})$.
Harmonic structures occupy the region $\mathfrak{u} = \mathcal{O}(1)$. That is, 
 $\frac{p \cdot k}{m^2} \frac{d \sigma^\mathrm{IPA}}{d u}$
serves as a unique master curve irrespective of the actual values of $u \ll 1$ or 
$p \cdot k/m^2$ or $a_0 \ll 1$.
For instance, keeping $p \cdot k/m^2$ and $a_0$ constant generates the master curve
in Fig.~\ref{fig.M}-left, $\frac{p \cdot k}{m^2} \frac{d \sigma^\mathrm{IPA}}{d u}$
vs.\ $\mathfrak{u} =u/u_1$, as a concise representation of harmonic structures around $u/u_1 \approx 1$,
as exhibited in Fig.~\ref{fig.a0dep} as a function of $u$.
Keeping $u$ and $a_0$ constant, the master curve in Fig.~\ref{fig.M}-right for the harmonic structures
in Fig.~\ref{fig.pkm2dep} as a function of $p \cdot k/m^2$
is generated; here $\mathfrak{u}^{-1} =  \frac{p \cdot k}{m^2} \frac{2}{u a_0^2}$ serves a
bottom axis. 
The extension to a master curve as a function of $a_0$ scaled by $\sqrt{u m^2/2 p \cdot k}$
is obvious. The representation as a function of $a_0^2 \frac{u m^2}{p \cdot k}$ reproduces Fig.~\ref{fig.M}-left.

When $a_0 < 1$, the second and higher harmonics do not form local maxima; instead, the $u$ spectra
become down stair-like. (Pulses with envelope function $g(\phi) = 1/\cosh (\phi /10 \pi)$ exhibit a similar
-- but further smoothed -- spectrum, cf.\ Fig.~2 in \cite{Acosta:2021iyu}.)
Analogously, larger values of $u$, $u > 1$, facilitate structureless distributions as a function of $a_0$,
displaying the ``rise and fall" behavior as in Fig.~3 in \cite{Acosta:2021iyu}, where also
pulses are considered for $a_0 \le 1$.
Reference \cite{King:2020hsk} emphasizes the appearance of an additional mid-IR peak, related to interference and
$n = 0$ contributions in pulses for $a_0 = \mathcal{O}(1)$.

\section{\boldmath $\omega'$ \unboldmath and \boldmath $\Theta'$ \unboldmath
differential distributions}\label{app.C}

After the concise consideration of the invariant differential cross section $d \sigma^\mathrm{IPA} / du$
and its dependencies, we turn to the differential cross section $d \sigma^\mathrm{IPA} / d \omega'$,
where $\omega' = \nu' m$ is the energy (scaled frequency) of the $out$ photon.
One cannot directly employ $d \sigma^\mathrm{IPA} / du$,
Eq.~(\ref{eq.5}), to get $d \sigma^\mathrm{IPA} / d \omega'$,
instead one must use a differently weighted summation, in general, 
\begin{align}
\frac{d \sigma^\mathrm{IPA}}{d \omega'} = - \frac{2 \pi \alpha^2}{p \cdot k}
\sum_{n=1}^{N \to \infty} \Theta(u_n - u(n,\nu') ) \frac{1}{m \kappa_n} \mathcal{F}_n (\hat z_n) , 
\end{align}
where $\hat z_n$ refers to the definition Eq.~(\ref{eq.7}) with $u \to u(n, \nu') := \frac{n \nu - \nu'}{\kappa_n - n \nu + \nu'}$,
see \cite{Kampfer:2020cbx,Harvey:2009ry} for details.
The quantity $\kappa_n := n \nu - \frac12 e^\zeta + \frac12 (1 + a_0^2) e^{- \zeta}$
with electron rapidity $\zeta = \mathrm{Arcosh} E_{e^-} / m$ can be positive or negative,
meaning read or blue shift. Here, $\nu := \omega / m$.

\begin{figure}[b]
\includegraphics[width=0.77\linewidth]{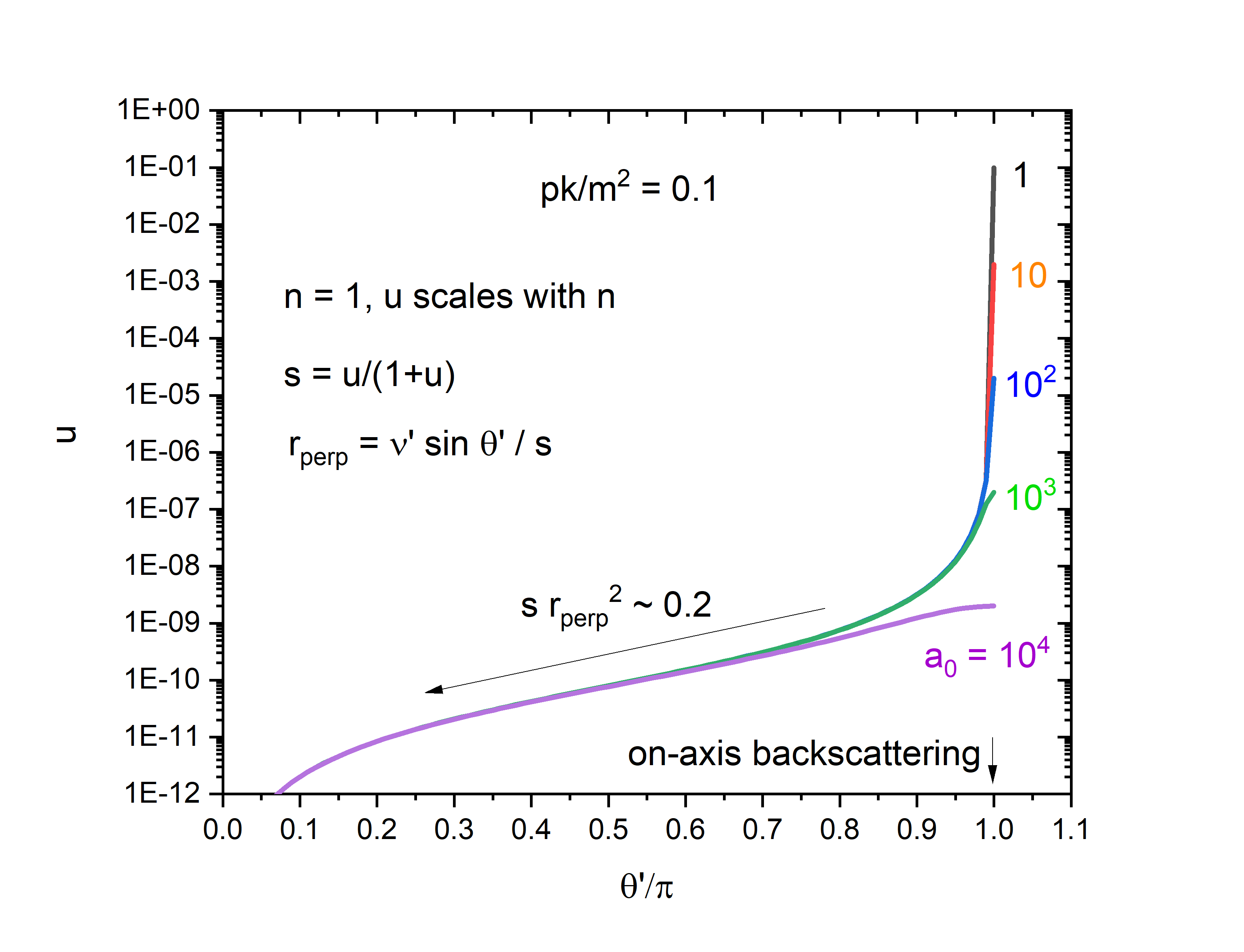} \\[-3mm]
\caption{The quantity $u(n=1, \Theta')$ as a function of the photon scattering angle $\Theta'$
for various values of $a_0 = 1, \cdots, 10^4$. 
The r.h.s.\ endpoints are at $u(n=1, \Theta' = \pi) = u_1$ (cf.\ Eq.~(\ref{eq.8})).
To accomplish larger values of $u$ at given $\Theta'$, $n > 1$ is required, pointing towards the CCF regime.
In contrast, in the region where $u(n=1, \Theta') \approx u^\mathrm{KN}(\Theta')$, the differential cross section
is saturated by the first harmonic (cf.\ Fig.~\ref{fig.a0dep}) and
$d \sigma^\mathrm{IPA}/d u \approx  d \sigma^\mathrm{KN}/d u$ applies
(cf.\ Eq.~(\ref{eq.14})).
The chosen value $p \cdot k/m^2 = 0.1$ arises from 
$e^\zeta = 5 \times 10^4$ and $\nu = \omega / m = 2 \times 10^{-6}$.
Note the scaling $u(n, \Theta') = n u(1, \Theta')$.   
The faint squares are for the KN kinematics $u^\mathrm{KN}(\Theta') := \frac{2 \nu (1- \cos \Theta')}
{e^\zeta (1+\cos \Theta')+e^{- \zeta} (1-\cos \Theta')}$.
\label{fig.u_Theta}
}
\end{figure}

The differential cross section $d \sigma^\mathrm{IPA} / d \omega'$ depends now on four parameters:
$\omega'$ (for the emitted photon in the lab.) and $a_0$ (external field intensity)
and $\omega$ (frequency of the external field in lab.) and $E_{e^-}$ (electron initial energy in lab.),
while the invariant differential cross section $d \sigma^\mathrm{IPA} / du$ had only three parameters, 
$u$ and $a_0$ and $p \cdot k/m^2$,
with the latter one being a combination of $\omega$ and $E_{e^-}$ for head-on collisions.
That is why the invariant differential cross section allows for a better survey of dependencies.
For similarities and differences of both cross sections and peculiarities w.r.t.\ red shift vs.\ blue shifts,
cf.\  \cite{Harvey:2009ry}.
In an experiment, where both the frequency $\omega'$ and angle $\theta'$ of the $out$ photon
are individually registered in each scattering event,
the variable $u$ can be constructed by $u = \frac{e^{-\zeta} \nu' (1- \cos \theta')}{1 - e^{-\zeta}\nu' (1- \cos \theta')}$.
Without fixing the respective harmonic number $n$, the inversions $u \mapsto \omega'$ or $u \mapsto \theta'$
are not possible, in general.
Considering 
$u(n, \Theta') = \frac{2 n \nu e^\zeta (1 - \cos \Theta')}{e^{2 \zeta} (1 + \cos \Theta') + (1+a_0^2)(1-\cos \Theta')}$
\cite{Harvey:2009ry},
instead of $u(n, \omega')$, yields the pattern exhibited in Fig.~\ref{fig.u_Theta}.
Of course, no dynamics is included here; it is simply kinematics aimed at exposing the statements in
\cite{Ilderton:2018nws,Blackburn:2021rqm}: the small-$u$ region refers to large scattering angles too.
For the considered kinematics, $u$ of the first harmonic, $n=1$, is independent of the intensity parameter $a_0$
as long as $a_0 \le 10^4$ and $\Theta' < 0.8 \pi$ (see inclined arrow); for smaller $a_0$ this region extends towards
the backscattering region. In the displayed example, $\mathfrak{s} r_\perp^2 \approx 0.2$, where
$r_\perp := \frac{1}{\mathfrak{s}} \nu' \sin \Theta'$ is the scaled transverse momentum of the $out$ photon and, again, 
$\mathfrak{s} = \frac{u}{1+u}$.
Such relations are at work w.r.t.\ the coincidence of IPA with KN in the small-$u$ region, see Fig.~\ref{fig.a0dep},
also showing that restricting simulation codes to on-axis back scattering-only 
(vertical arrow) is not perfect,
and beaming \cite{Blackburn:2019lgk} must be included.

\end{appendix}

\begin{acknowledgments}
The authors gratefully acknowledge the former collaboration with 
D.~Seipt, T.~Nousch, T.~Heinzl, A.~Otto
and useful discussions with
A.~Ilderton, K.~Krajewska,  M.~Marklund, C.~M\"uller, S.~Rykovanov, and G.~Torgrimsson.
A.~Ringwald and B.~King are thanked for explanations w.r.t.\ LUXE.
The work is supported by R.~Sauerbrey and T.~E.~Cowan w.r.t.\ the study
of fundamental QED processes for HIBEF. The present documentation (being a survey from a particle physics perspective, thus our preference of cross sections)
has been initiated by the HZDR HIBEF group. 
Useful discussions with R.~Sch\"utzhold, C.~Kohlf\"urst, and N.~Ahmadiniaz
are thanked for. 

The work of UHA was partly funded by the Center for Advanced Systems Understanding (CASUS) that is financed by Germany’s Federal Ministry of Education and Research (BMBF) and by the Saxon Ministry for Science, Culture and Tourism (SMWK) with tax funds on the basis of the budget approved by the Saxon State Parliament.
\end{acknowledgments}

\section*{Data Availability Statement}
This manuscript has no associated data. The code used to produce the plots in this manuscript are available upon request from the corresponding author.

\section*{Author Contribution Statement}
All authors contributed equally to the preparation of this manuscript.
 
{}


\begin{thebibliography}{99} 

\bibitem{Klein:1929lcc}
O.~Klein and Y.~Nishina,
``\"Uber die Streuung von Strahlung durch freie Elektronen nach der neuen relativistischen Quantendynamik von Dirac,''
Z. Phys. \textbf{52}, no.11, 853-868 (1929)

\bibitem{Fang:2020dmi}
K.~Fang, X.~J.~Bi, S.~J.~Lin and Q.~Yuan,
``Klein\textendash{}Nishina Effect and the Cosmic Ray Electron Spectrum,''
Chin. Phys. Lett. \textbf{38}, no.3, 039801 (2021)
[arXiv:2007.15601 [astro-ph.HE]].

\bibitem{Schlickeiser:2009qq}
R.~Schlickeiser and J.~Ruppel,
``Klein-Nishina steps in the energy spectrum of galactic cosmic ray electrons,''
New J. Phys. \textbf{12}, 033044 (2010)
[arXiv:0908.2183 [astro-ph.HE]].

\bibitem{LHAASO:2023gne}
Z.~Cao \textit{et al.} [LHAASO],
``Measurement of Ultra-High-Energy Diffuse Gamma-Ray Emission of the Galactic Plane from 10~TeV to 1~PeV with LHAASO-KM2A,''
Phys. Rev. Lett. \textbf{131}, no.15, 15 (2023)
[arXiv:2305.05372 [astro-ph.HE]].

\bibitem{LHAASO:2021gok}
Z.~Cao \textit{et al.} [LHAASO],
``Ultrahigh-energy photons up to 1.4 petaelectronvolts from 12 $\gamma$-ray Galactic sources,''
Nature \textbf{594}, no.7861, 33-36 (2021)
doi:10.1038/s41586-021-03498-z

\bibitem{Yakimenko:2018kih}
V.~Yakimenko, S.~Meuren, F.~Del Gaudio, C.~Baumann, A.~Fedotov, F.~Fiuza, T.~Grismayer, M.~J.~Hogan, A.~Pukhov and L.~O.~Silva, \textit{et al.}
``Prospect of Studying Nonperturbative QED with Beam-Beam Collisions,''
Phys. Rev. Lett. \textbf{122}, no.19, 190404 (2019)
[arXiv:1807.09271 [physics.plasm-ph]].

\bibitem{Abramowicz:2021zja}
H.~Abramowicz, U.~Acosta, M.~Altarelli, R.~A\ss{}mann, Z.~Bai, T.~Behnke, Y.~Benhammou, T.~Blackburn, S.~Boogert and O.~Borysov, \textit{et al.}
``Conceptual design report for the LUXE experiment,''
Eur. Phys. J. ST \textbf{230}, no.11, 2445-2560 (2021)
[arXiv:2102.02032 [hep-ex]].

\bibitem{Salgado:2021fgt}
F.~C.~Salgado, N.~Cavanagh, M.~Tamburini, D.~W.~Storey, R.~Beyer, P.~H.~Bucksbaum, Z.~Chen, A.~Di Piazza, E.~Gerstmayr and Harsh, \textit{et al.}
``Single particle detection system for strong-field QED experiments,''
New J. Phys. \textbf{24}, no.1, 015002 (2022)
[arXiv:2107.03697 [hep-ex]].


\bibitem{Meuren:2020nbw}
S.~Meuren, P.~H.~Bucksbaum, N.~J.~Fisch, F.~Fi\'uza, S.~Glenzer, M.~J.~Hogan, K.~Qu, D.~A.~Reis, G.~White and V.~Yakimenko,
[arXiv:2002.10051 [physics.plasm-ph]].


\bibitem{Storey:2023wlu}
D.~Storey, E.~Adli, J.~Allen, L.~Alsberg, R.~Ariniello, L.~Berman, P.~Bucksbaum, J.~Cao, C.~Clarke and S.~Corde, \textit{et al.}
``Status and first results from FACET-II towards the demonstration of plasma wakefield acceleration, coherent radiation generation, and probing strong-field QED,''
JACoW \textbf{IPAC2023}, TUPA104 (2023)

\bibitem{Brodin:2022dkd}
G.~Brodin, H.~Al-Naseri, J.~Zamanian, G.~Torgrimsson and B.~Eliasson,
``Plasma dynamics at the Schwinger limit and beyond,''
Phys. Rev. E \textbf{107}, no.3, 035204 (2023)
[arXiv:2209.07872 [physics.plasm-ph]].

\bibitem{Seipt:2023bcw}
D.~Seipt and A.~G.~R.~Thomas,
``Kinetic theory for spin-polarized relativistic plasmas,''
Phys. Plasmas \textbf{30}, no.9, 093102 (2023)
[arXiv:2307.02114 [physics.plasm-ph]].

\bibitem{Zhang:2020lxl}
P.~Zhang, S.~S.~Bulanov, D.~Seipt, A.~V.~Arefiev and A.~G.~R.~Thomas,
``Relativistic Plasma Physics in Supercritical Fields,''
Phys. Plasmas \textbf{27}, no.5, 050601 (2020)
[arXiv:2001.00957 [physics.plasm-ph]].

\bibitem{Poder:2017dpw}
K.~Poder, M.~Tamburini, G.~Sarri, A.~Di Piazza, S.~Kuschel, C.~D.~Baird, K.~Behm, S.~Bohlen, J.~M.~Cole and D.~J.~Corvan, \textit{et al.}
``Experimental Signatures of the Quantum Nature of Radiation Reaction in the Field of an Ultraintense Laser,''
Phys. Rev. X \textbf{8}, no.3, 031004 (2018)
[arXiv:1709.01861 [physics.plasm-ph]].

\bibitem{Cole:2017zca}
J.~M.~Cole, K.~T.~Behm, E.~Gerstmayr, T.~G.~Blackburn, J.~C.~Wood, C.~D.~Baird, M.~J.~Duff, C.~Harvey, A.~Ilderton and A.~S.~Joglekar, \textit{et al.}
``Experimental evidence of radiation reaction in the collision of a high-intensity laser pulse with a laser-wakefield accelerated electron beam,''
Phys. Rev. X \textbf{8}, no.1, 011020 (2018)
[arXiv:1707.06821 [physics.plasm-ph]].

\bibitem{Blinne:2018nbd}
A.~Blinne, H.~Gies, F.~Karbstein, C.~Kohlf\"urst and M.~Zepf,
``All-optical signatures of quantum vacuum nonlinearities in generic laser fields,''
Phys. Rev. D \textbf{99}, no.1, 016006 (2019)
[arXiv:1811.08895 [physics.optics]].

\bibitem{Gies:2017ezf}
H.~Gies, F.~Karbstein, C.~Kohlf\"urst and N.~Seegert,
``Photon-photon scattering at the high-intensity frontier,''
Phys. Rev. D \textbf{97}, no.7, 076002 (2018)
[arXiv:1712.06450 [hep-ph]].

\bibitem{Gies:2017ygp}
H.~Gies, F.~Karbstein and C.~Kohlf\"urst,
``All-optical signatures of Strong-Field QED in the vacuum emission picture,''
Phys. Rev. D \textbf{97}, no.3, 036022 (2018)
[arXiv:1712.03232 [hep-ph]].

\bibitem{Fedotov:2022ely}
A.~Fedotov, A.~Ilderton, F.~Karbstein, B.~King, D.~Seipt, H.~Taya and G.~Torgrimsson,
``Advances in QED with intense background fields,''
Phys. Rept. \textbf{1010}, 1-138 (2023)
[arXiv:2203.00019 [hep-ph]].

\bibitem{HernandezAcosta:2023msl}
U.~Hernandez Acosta and B.~K\"ampfer,
``Strong-field QED in the Furry-picture momentum-space formulation: Ward identities and Feynman diagrams,''
Phys. Rev. D \textbf{108}, no.1, 016013 (2023)
[arXiv:2303.12941 [hep-ph]].

\bibitem{Gonoskov:2021hwf}
A.~Gonoskov, T.~G.~Blackburn, M.~Marklund and S.~S.~Bulanov,
``Charged particle motion and radiation in strong electromagnetic fields,''
Rev. Mod. Phys. \textbf{94}, no.4, 045001 (2022)
[arXiv:2107.02161 [physics.plasm-ph]].

\bibitem{DiPiazza:2011tq}
A.~Di Piazza, C.~Muller, K.~Z.~Hatsagortsyan and C.~H.~Keitel,
``Extremely high-intensity laser interactions with fundamental quantum systems,''
Rev. Mod. Phys. \textbf{84}, 1177 (2012)
[arXiv:1111.3886 [hep-ph]].

\bibitem{Titov:2015pre}
A.~I.~Titov, B.~Kampfer, A.~Hosaka and H.~Takabe,
``Quantum processes in short and intensive electromagnetic fields,''
Phys. Part. Nucl. \textbf{47}, no.3, 456-487 (2016)
[arXiv:1512.07987 [hep-ph]].

\bibitem{Kaminski:2009wwd}
J.~Z.~Kami\'nski, K.~Krajewska and F.~Ehlotzky,
``Fundamental processes of quantum electrodynamics in laser fields of relativistic power,''
Rept. Prog. Phys. \textbf{72}, no.4, 046401 (2009)

\bibitem{Ivanov:2004fi}
D.~Y.~Ivanov, G.~L.~Kotkin and V.~G.~Serbo,
Eur. Phys. J. C \textbf{36}, 127-145 (2004)
doi:10.1140/epjc/s2004-01861-x
[arXiv:hep-ph/0402139 [hep-ph]].

\bibitem{Ritus85}
  V. I. Ritus,
 ``Quantum effects of the interaction of elementary particles
with an intense electromagnetic field”.
 J. Sov. Laser Res. (United States), \textbf{6:5}, 497 (1985).

\bibitem{Gelfer:2022nqy}
E.~G.~Gelfer, A.~M.~Fedotov, A.~A.~Mironov and S.~Weber,
``Nonlinear Compton scattering in time-dependent electric fields beyond the locally constant crossed field approximation,''
Phys. Rev. D \textbf{106}, no.5, 056013 (2022)
[arXiv:2206.08211 [hep-ph]].

\bibitem{Ilderton:2018nws}
A.~Ilderton, B.~King and D.~Seipt,
``Extended locally constant field approximation for nonlinear Compton scattering,''
Phys. Rev. A \textbf{99}, no.4, 042121 (2019)
[arXiv:1808.10339 [hep-ph]].

\bibitem{Nielsen:2021nuo}
C.~F.~Nielsen, R.~Holtzapple and B.~King,
``High-resolution modeling of nonlinear Compton scattering in focused laser pulses,''
Phys. Rev. D \textbf{106}, no.1, 013010 (2022)
[arXiv:2109.09490 [physics.plasm-ph]].

\bibitem{Acosta:2021iyu}
U.~Hernandez~Acosta, A.~I.~Titov and B.~K\"ampfer,
``Rise and fall of laser-intensity effects in spectrally resolved Compton process,''
New J. Phys. \textbf{23}, no.9, 095008 (2021)
[arXiv:2105.11758 [hep-ph]].

\bibitem{Kampfer:2020cbx}
B.~K\"ampfer and A.~I.~Titov,
``Impact of laser polarization on q-exponential photon tails in non-linear Compton scattering,''
Phys. Rev. A \textbf{103}, 033101 (2021)
[arXiv:2012.07699 [hep-ph]].

\bibitem{Mackenroth:2010jr}
F.~Mackenroth and A.~Di Piazza,
``Nonlinear Compton scattering in ultra-short laser pulses,''
Phys. Rev. A \textbf{83}, 032106 (2011)
[arXiv:1010.6251 [hep-ph]].

\bibitem{Heinzl:2009nd}
T.~Heinzl, D.~Seipt and B.~Kampfer,
``Beam-Shape Effects in Nonlinear Compton and Thomson Scattering,''
Phys. Rev. A \textbf{81}, 022125 (2010)
[arXiv:0911.1622 [hep-ph]].

\bibitem{Seipt:2010ya}
D.~Seipt and B.~Kampfer,
``Non-Linear Compton Scattering of Ultrashort and Ultraintense Laser Pulses,''
Phys. Rev. A \textbf{83}, 022101 (2011)
[arXiv:1010.3301 [hep-ph]].

\bibitem{Seipt:2011zz}
D.~Seipt and B.~Kampfer,
``Scaling law for the photon spectral density in the nonlinear Thomson-Compton scattering,''
Phys. Rev. ST Accel. Beams \textbf{14}, 040704 (2011)

\bibitem{Seipt:2011dx}
D.~Seipt and B.~Kampfer,
``Non-linear Compton scattering of ultrahigh-intensity laser pulses,''
Laser Phys. \textbf{23}, 075301 (2013)
[arXiv:1111.0188 [hep-ph]].

\bibitem{Seipt:2013taa}
D.~Seipt and B.~Kampfer,
``Asymmetries of azimuthal photon distributions in non-linear Compton scattering in ultra-short intense laser pulses,''
Phys. Rev. A \textbf{88}, 012127 (2013)
[arXiv:1305.3837 [physics.optics]].

\bibitem{Ilderton:2020dhs}
A.~Ilderton, B.~King and S.~Tang,
``Toward the observation of interference effects in nonlinear Compton scattering,''
Phys. Lett. B \textbf{804}, 135410 (2020)
[arXiv:2002.04629 [physics.atom-ph]].

\bibitem{Seipt:2016rtk}
D.~Seipt, V.~Kharin, S.~Rykovanov, A.~Surzhykov and S.~Fritzsche,
``Analytical results for nonlinear Compton scattering in short intense laser pulses,''
J. Plasma Phys. \textbf{82}, no.2, 655820203 (2016)
[arXiv:1601.00442 [hep-ph]].

\bibitem{Dinu:2018efz}
V.~Dinu and G.~Torgrimsson,
``Single and double nonlinear Compton scattering,''
Phys. Rev. D \textbf{99}, no.9, 096018 (2019)
[arXiv:1811.00451 [hep-ph]].

\bibitem{Titov:2023yev}
A.~I.~Titov,
``Polarization of recoil photon in nonlinear Compton process,''
Eur. Phys. J. D \textbf{78}, no.3, 31 (2024)
[arXiv:2307.00620 [hep-ph]].

\bibitem{Titov:2015tdz}
A.~I.~Titov, B.~Kampfer, A.~Hosaka, T.~Nousch and D.~Seipt,
``Determination of the carrier envelope phase for short, circularly polarized laser pulses,''
Phys. Rev. D \textbf{93}, no.4, 045010 (2016)
[arXiv:1512.07504 [hep-ph]].

\bibitem{Hartin:2011vr}
A.~Hartin and G.~Moortgat-Pick,
``High Intensity Compton Scattering in a strong plane wave field of general form,''
Eur. Phys. J. C \textbf{71}, 1729 (2011)
[arXiv:1106.1671 [hep-th]].

\bibitem{Angioi:2016vir}
A.~Angioi, F.~Mackenroth and A.~Di Piazza,
``Nonlinear single Compton scattering of an electron wave-packet,''
Phys. Rev. A \textbf{93}, no.5, 052102 (2016)
[arXiv:1602.02639 [hep-ph]].

\bibitem{Blackburn:2018sfn}
T.~G.~Blackburn, D.~Seipt, S.~S.~Bulanov and M.~Marklund,
``Benchmarking semiclassical approaches to strong-field QED: nonlinear Compton scattering in intense laser pulses,''
Phys. Plasmas \textbf{25}, no.8, 083108 (2018)
[arXiv:1804.11085 [physics.plasm-ph]].

\bibitem{Lotstedt:2009zz}
E.~Lotstedt and U.~D.~Jentschura,
``Nonperturbative Treatment of Double Compton Backscattering in Intense Laser Fields,''
Phys. Rev. Lett. \textbf{103}, 110404 (2009)
[arXiv:0909.4984 [quant-ph]].

\bibitem{Guccione-Gush:1975juz}
R.~Guccione-Gush and H.~P.~Gush,
``Photon Emission by an electron in a bichromatic Field,''
Phys. Rev. D \textbf{12}, 404-414 (1975)

\bibitem{Herrmann:1971jc}
J.~Herrmann and V.~C.~Zhukovskii,
``Compton scattering and induced Compton scattering in a constant electromagnetic field,''
Annalen Phys. \textbf{27}, 349-358 (1971)

\bibitem{Ehlotzky:1989}
F.~Ehlotzky,
``Scattering of x-rays by relativistic electrons in a strong laser field,"
J.~Phys.~B \textbf{22}, 601 (1989)

\bibitem{Akhiezer:1985}
A.~I.~Akhiezer and N.~P.~Merenov,
``Scattering of a photon by an electron moving in the field of a plane periodic electromagnetic wave,"
Sov.~Phys.~JETP \textbf{61}, 1 (1985)

\bibitem{Mackenroth:2012rb}
F.~Mackenroth and A.~Di Piazza,
``Nonlinear Double Compton Scattering in the Ultrarelativistic Quantum Regime,''
Phys. Rev. Lett. \textbf{110}, no.7, 070402 (2013)
[arXiv:1208.3424 [hep-ph]].

\bibitem{Seipt:2012tn}
D.~Seipt and B.~K\"ampfer,
``Two-photon Compton process in pulsed intense laser fields,''
Phys. Rev. D \textbf{85}, 101701 (2012)
[arXiv:1201.4045 [hep-ph]].

\bibitem{Seipt:2013hda}
D.~Seipt and B.~K\"ampfer,
``Laser assisted Compton scattering of X-ray photons,''
Phys. Rev. A \textbf{89}, no.2, 023433 (2014)
[arXiv:1309.2092 [physics.atom-ph]].

\bibitem{Lotstedt:2013uya}
E.~L\"otstedt and U.~D.~Jentschura,
``Theoretical study of the Compton effect with correlated three-photon emission: From the differential cross section to high-energy triple-photon entanglement,''
Phys. Rev. A \textbf{87}, no.3, 033401 (2013)
[arXiv:1405.1669 [quant-ph]].

\bibitem{Lotstedt:2012zz}
E.~Lotstedt and U.~D.~Jentschura,
``Triple Compton Effect: A Photon Splitting into Three upon Collision with a Free Electron,''
Phys. Rev. Lett. \textbf{108}, 233201 (2012)
[arXiv:1205.0317 [quant-ph]].

\bibitem{Dinu:2019pau}
V.~Dinu and G.~Torgrimsson,
``Approximating higher-order nonlinear QED processes with first-order building blocks,''
Phys. Rev. D \textbf{102}, no.1, 016018 (2020)
[arXiv:1912.11015 [hep-ph]].

\bibitem{Sidi:2003bnm}
A.~Sidi,
``Practical extrapolation methods: Theory and applications,''
Cambridge University Press, Vol. 10. (2003).

\bibitem{Ritus:1972ky}
V.~I.~Ritus,
``Radiative corrections in quantum electrodynamics with intense field and their analytical properties,''
Annals Phys. \textbf{69}, 555-582 (1972)

\bibitem{Fedotov:2016afw}
A.~M.~Fedotov,
``Conjecture of perturbative QED breakdown at $\alpha\chi^{2/3} \gtrsim 1$,''
J. Phys. Conf. Ser. \textbf{826}, no.1, 012027 (2017)
[arXiv:1608.02261 [hep-ph]].

\bibitem{Titov:2019kdk}
A.~I.~Titov, A.~Otto and B.~Kampfer,
``Multi-photon regime of non-linear Breit-Wheeler and Compton processes in short linearly and circularly polarized laser pulses,''
Eur. Phys. J. D \textbf{74}, no.2, 39 (2020)
[arXiv:1907.00643 [physics.plasm-ph]].

\bibitem{Heinzl:2021mji}
T.~Heinzl, A.~Ilderton and B.~King,
``Classical Resummation and Breakdown of Strong-Field QED,''
Phys. Rev. Lett. \textbf{127}, no.6, 061601 (2021)
[arXiv:2101.12111 [hep-ph]].

\bibitem{Mironov:2020gbi}
A.~A.~Mironov, S.~Meuren and A.~M.~Fedotov,
``Resummation of QED radiative corrections in a strong constant crossed field,''
Phys. Rev. D \textbf{102}, no.5, 053005 (2020)
[arXiv:2003.06909 [hep-th]].

\bibitem{Blackburn:2021cuq}
T.~G.~Blackburn and B.~King,
``Higher fidelity simulations of nonlinear Breit\textendash{}Wheeler pair creation in intense laser pulses,''
Eur. Phys. J. C \textbf{82}, no.1, 44 (2022)
[arXiv:2108.10883 [hep-ph]].

\bibitem{Heinzl:2020ynb}
T.~Heinzl, B.~King and A.~J.~Macleod,
``The locally monochromatic approximation to QED in intense laser fields,''
Phys. Rev. A \textbf{102}, 063110 (2020)
[arXiv:2004.13035 [hep-ph]].

\bibitem{DiPiazza:2018bfu}
A.~Di Piazza, M.~Tamburini, S.~Meuren and C.~H.~Keitel,
``Improved local-constant-field approximation for strong-field QED codes,''
Phys. Rev. A \textbf{99}, no.2, 022125 (2019)
[arXiv:1811.05834 [hep-ph]].

\bibitem{DiPiazza:2017raw}
A.~Di Piazza, M.~Tamburini, S.~Meuren and C.~H.~Keitel,
``Implementing nonlinear Compton scattering beyond the local constant field approximation,''
Phys. Rev. A \textbf{98}, no.1, 012134 (2018)
[arXiv:1708.08276 [hep-ph]].

\bibitem{LLIV} W. B. Berestetzki, E. M. Lifschitz, and L. P. Pitajewski, Relativistische Qantentheorie, 
Lehrbuch der Theoretischen Physik, Vol. IV (Akademie Verlag, Berlin, 1980). 

\bibitem{Lee:2021iid}
R.~N.~Lee, M.~D.~Schwartz and X.~Zhang,
``Compton Scattering Total Cross Section at Next-to-Leading Order,''
Phys. Rev. Lett. \textbf{126}, no.21, 211801 (2021)
[arXiv:2102.06718 [hep-ph]].

\bibitem{Bragin:2020akq}
S.~Bragin and A.~Di Piazza,
``Electron-positron annihilation into two photons in an intense plane-wave field,''
Phys. Rev. D \textbf{102}, no.11, 116012 (2020)
[arXiv:2003.02231 [hep-ph]].

\bibitem{King:2014wfa}
B.~King,
``Double Compton scattering in a constant crossed field,''
Phys. Rev. A \textbf{91}, no.3, 033415 (2015)
[arXiv:1410.5478 [hep-ph]].

\bibitem{Narozhnyi:1980dc}
N.~B.~Narozhnyi,
``Expansion parameter of perturbation theory in intense field quantum electrodynamics,"
Phys. Rev. D \textbf{21}, 1176-1183 (1980)

\bibitem{Podszus:2018hnz}
T.~Podszus and A.~Di Piazza,
``High-energy behavior of strong-field QED in an intense plane wave,''
Phys. Rev. D \textbf{99}, no.7, 076004 (2019)
[arXiv:1812.08673 [hep-ph]].

\bibitem{Ilderton:2019kqp}
A.~Ilderton,
``Note on the conjectured breakdown of QED perturbation theory in strong fields,''
Phys. Rev. D \textbf{99}, no.8, 085002 (2019)
[arXiv:1901.00317 [hep-ph]].

\bibitem{Lotstedt:2008}
E.~L\"otstedt, U.~D.~Jentschura, and C.~H.~Keitel, 
``Laser channeling of Bethe-Heitler pairs,"
Phys. Rev. Lett. 101, 203001 (2008)

\bibitem{Khokonov:2005}
Khokonov, A. Kh, and M. Kh Khokonov,
``Classification of the interactions of relativistic electrons with laser radiation,"
Tech. Phys. Lett. \textbf{31}, 154-156 (2005)

\bibitem{HIBEF}
Helmholtz International Beam Line for Extreme Fields (HIBEF),
\begin{verbatim} https://www.hzdr.de/db/Cms?pOid=50566&pNid=694 \end{verbatim}

\bibitem{Ahmadiniaz:2024xob}
N.~Ahmadiniaz, C.~B\"ahtz, A.~Benediktovitch, C.~B\"omer, L.~Bocklage, T.~E.~Cowan, J.~Edwards, S.~Evans, S.~F.~Vi\~nas and H.~Gies, \textit{et al.}
``Letter of Intent: Towards a Vacuum Birefringence Experiment at the Helmholtz International Beamline for Extreme Fields,''
[arXiv:2405.18063 [physics.ins-det]].

\bibitem{ZEUS}
Zettawatt-Equivalent Ultrashort pulse laser System (NFS ZEUS), \url{https://zeus.engin.umich.edu }, Gérard Mourou Center for Ultrafast Optical Science, 	Ann Arbor (MI)

\bibitem{OPAL}
Optical Parametric Amplifier Line (NSF OPAL), \url{https://nsf-opal.rochester.edu}, University of Rochester, Rochester (NY)

\bibitem{SELabc}
SEL-100 PW laser facility, \url{http://english.siom.cas.cn/Newsroom/rp/202207/t20220701_307101.html}, Shanghai High repetition rate XFEL and Extreme light facility (SHINE), Shanghai

\bibitem{RELAX}
A.~Garcia Laso, H.~H{\"o}ppner, A.~Pelka, C.~B{\"a}htz, S.~Di Dio Cafiso, J. Dreyer, M.~Hassan, T.~Kluge \textit{et al.},
``ReLaX: the Helmholtz International Beamline for Extreme Fields high-intensity short-pulse laser driver for relativistic laser-matter interaction and strong-field science using the high energy density instrument at the European X-ray free electron laser facility'',
High Power Laser Sci. Eng. \textbf{9}, e59 (2021)

\bibitem{EuXFEL}
U.~Zastrau \textit{et al.},
 ``The High Energy Density Scientific Instrument at the European XFEL'',
J. Synchrotron Radiat. 28, 1393 (2021)


\bibitem{Pentia:2023jqf}
M.~Pentia, C.~R.~Badita, D.~Dumitriu, A.~R.~Ionescu and H.~Petrascu,
``The Strong Field QED approach of the vacuum interaction processes at ELI-NP,''
[arXiv:2307.09315 [hep-ph]].

\bibitem{Gales:2018}
S.~Gales \textit{et al.},
``The extreme light infrastructure—nuclear physics (ELI-NP) facility: new horizons in physics with 10 PW ultra-intense lasers and 20 MeV brilliant gamma beams.'' 
Reports on Progress in Physics 81.9, 094301  (2018)


\bibitem{Yoon:2021ony}
J.~ W.~Yoon, Y.~G.~Kim, I.~W.~Choi, J.~H.~Sung, H.~W.~Lee, S.~K.~Lee, and C.~H.~Nam 
``Realization of laser intensity over $10^{23}$ \,W/cm${}^2$,''
Optica \textbf{8}, no.5, 630-635 (2021)

\bibitem{Boncioli:2023gbl}
D.~Boncioli,
``Cosmic-ray propagation in extragalactic space and secondary messengers,''
Proc. Int. Sch. Phys. Fermi \textbf{208}, 315-351 (2024)
[arXiv:2309.12743 [astro-ph.HE]].

\bibitem{HESS:2023sxo}
F.~Aharonian \textit{et al.} [H.E.S.S.],
``Discovery of a radiation component from the Vela pulsar reaching 20 teraelectronvolts,''
Nature Astron. \textbf{7}, no.11, 1341-1350 (2023)
[erratum: Nature Astron. \textbf{8}, no.1, 145 (2024)]
[arXiv:2310.06181 [astro-ph.HE]].

\bibitem{Li:2024oda}
L.~Li, S.~Q.~Zhong, D.~Xiao, Z.~G.~Dai, S.~F.~Huang and Z.~F.~Sheng,
``Magnetar as the Central Engine of AT2018cow: Optical, Soft X-Ray, and Hard X-Ray Emission,''
[arXiv:2402.15067 [astro-ph.HE]].

\bibitem{ParticleDataGroup:2022pth}
R.~L.~Workman \textit{et al.} [Particle Data Group],
``Review of Particle Physics,''
PTEP \textbf{2022}, 083C01 (2022)

\bibitem{Heinzl:2008rh}
T.~Heinzl and A.~Ilderton,
``A Lorentz and gauge invariant measure of laser intensity,''
Opt. Commun. \textbf{282}, 1879-1883 (2009)
[arXiv:0807.1841 [physics.class-ph]].

\bibitem{Harvey:2009ry}
C.~Harvey, T.~Heinzl and A.~Ilderton,
``Signatures of High-Intensity Compton Scattering,''
Phys. Rev. A \textbf{79}, 063407 (2009)
[arXiv:0903.4151 [hep-ph]].

\bibitem{Chen:2022dgo}
Y.~Y.~Chen, K.~Z.~Hatsagortsyan, C.~H.~Keitel and R.~Shaisultanov,
``Electron spin- and photon polarization-resolved probabilities of strong-field QED processes,''
Phys. Rev. D \textbf{105}, no.11, 116013 (2022)
[arXiv:2201.10863 [hep-ph]].

\bibitem{Gong:2022qis}
Z.~Gong, K.~Z.~Hatsagortsyan and C.~H.~Keitel,
``Electron Polarization in Ultrarelativistic Plasma Current Filamentation Instabilities,''
Phys. Rev. Lett. \textbf{130}, no.1, 015101 (2023)
[arXiv:2212.03303 [physics.plasm-ph]].

\bibitem{Song:2024oak}
H.~H.~Song and M.~Tamburini,
``Polarized QED cascades over pulsar polar caps,''
Mon. Not. Roy. Astron. Soc. \textbf{530}, no.2, 2087-2095 (2024)
[arXiv:2401.09829 [astro-ph.HE]].

\bibitem{Blackburn:2021rqm}
T.~G.~Blackburn, A.~J.~MacLeod and B.~King,
``From local to nonlocal: higher fidelity simulations of photon emission in intense laser pulses,''
New J. Phys. \textbf{23}, no.8, 085008 (2021)
[arXiv:2103.06673 [hep-ph]].

\bibitem{Blackburn:2019lgk}
T.~G.~Blackburn, D.~Seipt, S.~S.~Bulanov and M.~Marklund,
``Radiation beaming in the quantum regime,''
Phys. Rev. A \textbf{101}, no.1, 012505 (2020)
[arXiv:1904.07745 [physics.plasm-ph]].

\bibitem{King:2020hsk}
B.~King,
``Interference effects in nonlinear Compton scattering due to pulse envelope,''
Phys. Rev. D \textbf{103}, no.3, 036018 (2021)
[arXiv:2012.05920 [hep-ph]].

\bibitem{Titov:2024aez}
A.~I.~Titov,
``Effects of Photon Polarizations in Non-Linear Compton Process,''
Phys. Part. Nucl. \textbf{55}, no.4, 920-928 (2024)
doi:10.1134/S1063779624700588


\end{thebibliography}
\end{document}